\documentclass{article}

\usepackage{amssymb}
\usepackage{booktabs}
\usepackage{graphicx}
\usepackage{microtype}
\usepackage{subfigure}
\usepackage{pifont}
\usepackage[dvipsnames]{xcolor}
\usepackage{multirow}
\usepackage{siunitx}
\usepackage[aboveskip=0pt,belowskip=0pt]{caption}

\usepackage{hyperref} %
\usepackage{cleveref} %

\usepackage[accepted]{mlsys2025}

\usepackage[frozencache,cachedir=minted-cache]{minted}
\setminted{fontsize=\scriptsize}
\setminted{fontfamily=cmtt}
\usepackage{etoolbox,xpatch}
\makeatletter
\AtBeginEnvironment{minted}{\dontdofcolorbox}
\def\dontdofcolorbox{\renewcommand\fcolorbox[4][]{##4}}
\xpatchcmd{\inputminted}{\minted@fvset}{\minted@fvset\dontdofcolorbox}{}{}
\xpatchcmd{\mintinline}{\minted@fvset}{\minted@fvset\dontdofcolorbox}{}{}
\makeatother

\newcommand{\cmark}{\ding{51}}
\newcommand{\xmark}{\ding{55}}

\AddToHook{cmd/appendix/before}{%
  \crefalias{section}{appendix}%
  \crefalias{subsection}{appendix}%
}

\def\mytitle{\texttt{fabric-lib}: RDMA Point-to-Point Communication for LLM Systems}

\mlsystitlerunning{\mytitle}

\begin{document}

\twocolumn[
\mlsystitle{\mytitle}

\mlsyssetsymbol{equal}{*}

\begin{mlsysauthorlist}
\mlsysauthor{Nandor Licker}{equal}
\mlsysauthor{Kevin Hu}{}
\mlsysauthor{Vladimir Zaytsev}{}
\mlsysauthor{Lequn Chen}{equal}
\end{mlsysauthorlist}

\mlsyscorrespondingauthor{Nandor Licker}{nandor@perplexity.ai}
\mlsyscorrespondingauthor{Lequn Chen}{lequn@perplexity.ai}

\vskip 0.3in

\begin{abstract}
Emerging Large Language Model (LLM) system patterns, such as disaggregated inference, Mixture-of-Experts (MoE) routing, and asynchronous reinforcement fine-tuning, require flexible point-to-point communication beyond simple collectives.
Existing implementations are locked to specific Network Interface Controllers (NICs), hindering integration into inference engines and portability across hardware providers.
We present \texttt{fabric-lib}, which bridges the functionality of common NICs to expose a uniform interface.
\texttt{fabric-lib} exposes one-sided \textsc{WriteImm} operations with a \textsc{ImmCounter} primitive for completion notification, without ordering assumptions of network transport, transparently managing multiple NICs per GPU.
We demonstrate peak throughput of 400 Gbps on both NVIDIA ConnectX-7 and AWS Elastic Fabric Adapter (EFA).
We showcase \texttt{fabric-lib} through three production systems: (1) KvCache transfer for disaggregated inference with dynamic scaling, (2) RL weight updates achieving 1.3 seconds for trillion-parameter models, and (3) MoE dispatch/combine  implementation exceeding DeepEP decode latency on ConnectX-7, with the first viable latencies on EFA. We demonstrate that our portable point-to-point communication complements collectives while avoiding lock-in.
\texttt{fabric-lib} is open-sourced at \url{https://github.com/perplexityai/pplx-garden/}.
\end{abstract}

]

\let\thefootnote\relax\footnotetext{
  Perplexity AI.
  \quad\textsuperscript{*}Equal contribution
  \quad{}Correspondence to: Lequn Chen \texttt{<lequn@perplexity.ai>}.
  \\
  \\
  \textit{Proceedings of the 9\textsuperscript{th} MLSys Conference},
  Bellevue, WA, USA, 2026.
  Copyright 2026 by the author(s).
}

\section{Introduction}

Mixture-of-Experts (MoE) architectures are becoming the dominant approach for scaling model capacity~\cite{DBLP:conf/iclr/ShazeerMMDLHD17},
while disaggregated inference is emerging as the standard for production serving.~\cite{splitwise,DistServe,Mooncake2025}
These new architectures rely on communication patterns that are fundamentally different from traditional collective-based parallelism.

LLM frameworks overwhelmingly rely on collective communication, often through \textit{NCCL} or \texttt{torch.dist\-ributed}.~\cite{NCCL, torch.distributed}
While collectives excel at static patterns such as tensor or data parallelism~\cite{Megatron-LM, Horovod, FSDP},
they impose constraints unsuitable for emerging workloads: fixed membership prevents dynamic scaling, synchronized initialization adds overhead, and uniform buffer sizes force dense communication even for sparse patterns.
While these libraries offer \textsc{Send} and \textsc{Recv} primitives for point-to-point communication, they often cannot be effectively composed to achieve viable latency.

High-performance computing has long used primitives (\textsc{Send}, \textsc{Recv}, \textsc{Write}) built on Remote Direct Memory Access (RDMA) for flexible low-latency high-bandwidth transfers~\cite{DBLP:conf/usenix/KaliaKA16},
yet such primitives are rarely available in LLM frameworks.
The key barrier is hardware diversity without uniform abstraction.
Cloud providers deploy different RDMA implementations: NVIDIA ConnectX NICs use traditional Reliable Connection (RC) transport with in-order delivery, while AWS Elastic Fabric Adapter (EFA) implements a proprietary Scalable Reliable Datagram (SRD) protocol~\cite{AWS-SRD} with out-of-order delivery.
Existing solutions suffer from vendor lock-in: \textit{DeepEP}~\cite{DeepEP} requires GPU-initiated RDMA (IBGDA)~\cite{GPUDirectAsync} provided exclusively by ConnectX, \textit{NVSHMEM}~\cite{NVSHMEM} exhibits severe performance degradation on EFA and recent libraries like \textit{Mooncake}~\cite{Mooncake2025} and \textit{NIXL}~\cite{NIXL} lack EFA support or remain preliminary.
Consequently, there are no viable cross-provider solutions for LLM inference.

We address portability by leveraging the common functionality across heterogeneous RDMA hardware.
Our key insight is that both ConnectX and EFA support reliable but unordered delivery: ConnectX RC can ignore ordering, while EFA SRD is inherently unordered.
We introduce \texttt{fabric-lib}, a portable RDMA communication library.
It provides two-sided \textsc{Send}/\textsc{Recv} and one-sided \textsc{WriteImm} operations with a novel \textsc{ImmCounter} primitive for completion notification that does not rely on message ordering.
It exposes a common interface to both ConnectX and EFA, avoiding vendor lock-in.
It transparently manages multiple NICs per GPU, essential for EFA where four 100 Gbps NICs must be aggregated to reach 400 Gbps.

We demonstrate \texttt{fabric-lib} through production systems addressing the shortcomings of collectives:

\textbf{KvCache transfer} (\Cref{sec:kvcache}): Disaggregated inference with unrestricted prefiller-decoder communication, enabling elastic scaling without synchronized initialization or fixed membership.
Production tested on EFA with full CUDA Graph support and low-latency layer-by-layer transfers.

\textbf{RL weight update} (\Cref{sec:rollout}): Point-to-point approach achieves 1.3-second updates for trillion-parameter models, over 100$\times$ faster than existing RL frameworks.~\cite{Nemo-RL-Optimizing-Weight-Transfer,BiaoHe-Optimizing-Weight-Sync-in-slime}
Utilizes full cluster bandwidth by one-sided RDMA \textsc{Write} directly from each training GPU to inference GPUs.
Pipelined execution overlaps H2D memcpy, weight preparation, and RDMA transfer.

\textbf{MoE dispatch/combine} (\Cref{sec:moe}): State-of-the-art decode latency on ConnectX-7, competitive with specialized \textit{DeepEP} kernels despite the use of a host proxy thread.
First viable implementation on EFA, relying on parallel token and route transfers to hide device-to-host and network latency.

The systems we present span diverse communication patterns, ranging from paged writes to bulk transfers and coordinated scatter operations, all production-deployed on heterogeneous hardware.
Our results demonstrate that point-to-point communication complements collectives for modern LLM workloads, and that portable abstractions can avoid vendor lock-in while retaining performance.

\section{Background and Related Work}

\subsection{Network Technology}

\paragraph{RDMA}
Remote Direct Memory Access (RDMA) is the high-throughput, low-latency backbone of modern ML systems.
Presently deployed NICs deliver 400 Gbps bandwidth at sub-$\mu$s latencies.
RDMA achieves its performance through a split design: control plane operations initialize the device and set up buffers with operating system involvement, while data plane operations (data transfer and completion polling) bypass the kernel, avoiding system call overheads.

\textbf{RDMA Operations\quad}
RDMA provides both two-sided and one-sided operations.
Two-sided \textsc{Send}/\textsc{Recv} operations first post a \textsc{Recv} with a registered memory region, with the sender issueing a \textsc{Send},
notifying the receiver via a completion queue.
One-sided \textsc{Write} operations copy local memory to a remote memory region without peer involvement, requiring the remote memory address and key (\textsc{RKey}).
\textsc{WriteImm} extends \textsc{Write} by delivering a 32-bit immediate value that also notifies the receiver via the completion queue.
While \textsc{Read} and atomic operations are available, their latencies are not suitable for our purposes.~\cite{DBLP:conf/usenix/KaliaKA16,DBLP:conf/nsdi/RedaCKP22}

\textbf{RDMA Transports\quad}
The RDMA specification defines three transport protocols: Reliable Connection (RC), Unreliable Connection (UC), and Unreliable Datagram (UD).
\Cref{table:rdma-protocols} summarizes their capabilities, highlighting \texttt{fabric-lib} as the common ground between them. 

\begin{table}[t]
\begin{center}
\begin{small}
\begin{tabular}{c|ccccc}
\multicolumn{1}{c}{}       & RC       & UC     & UD     & SRD      & \textbf{\texttt{fabric-lib}} \\
\hline\hline
Reliability                & \cmark   & \xmark & \xmark & \cmark   & {\color{OliveGreen}\cmark} \\
Ordering                   & \cmark   & \xmark & \xmark & \xmark   & {\color{RedOrange}\xmark}  \\
Connection                 & \cmark   & \cmark & \xmark & \xmark   & {\color{RedOrange}\xmark}  \\
\textsc{Send}/\textsc{Recv}& \cmark   & \cmark & MTU    & \cmark   & {\color{OliveGreen}\cmark}               \\
\textsc{WriteImm}          & \cmark   & \cmark & \xmark & \cmark   & {\color{OliveGreen}\cmark} \\
\textsc{Read}              & \cmark   & \xmark & \xmark & \cmark   & {\color{RedOrange}\xmark}  \\
Atomic                     & \cmark   & \xmark & \xmark & \cmark   & {\color{RedOrange}\xmark}  \\
\hline
\end{tabular}
\end{small}
\end{center}
\caption{Comparison of RDMA transport types}
\label{table:rdma-protocols}
\end{table}

\textbf{Cloud RDMA Adapters\quad}
Beyond the widely-deployed ConnectX series, major cloud providers are deploying their own solutions, such as AWS Elastic Fabric Adapter (EFA),
Alibaba Cloud eRDMA and Google Falcon~\cite{eRDMA,Falcon}.
While most are RC-compatible, EFA implements the proprietary SRD protocol, exposed via \texttt{libfabric}.~\cite{AWS-SRD,libfabric}
SRD is connectionless and provides reliable but unordered delivery.

\subsection{Programming Interface}

\textbf{Collectives\quad} ML frameworks predominantly rely on collective communication libraries (e.g., \texttt{torch.distributed}, NCCL, MPI) for inter-GPU data exchange.~\cite{torch.distributed, NCCL, mpi50}
While collective operations excel at structured data transfers,~\cite{Megatron-LM, Horovod, FSDP} their point-to-point capabilities are limited:
\textbf{(1) Fixed membership} hinders dynamic scaling by requiring advance knowledge of all participants;
\textbf{(2) Synchronized initialization} mandates global coordination to form ``worlds'', blocking independent peer connections;
\textbf{(3) Operation ordering} requires all participants to agree on operation sequence, necessitating extra synchronization even when NCCL supports concurrent receives;~\cite{Demystifying-NCCL}
\textbf{(4) Shape uniformity} constrains transfer sizes across participants, even for point-to-point.

\textbf{GPUDirect\quad}
GPUDirect RDMA enables RDMA NICs to directly access GPU memory over PCIe, eliminating intermediate CPU memory copies.~\cite{GPUDirectRDMA}
Transfers can be initiated either by the host, or if GPUDirect Async (IBGDA) is available, from the GPU itself to bypass CPU overheads~\cite{GPUDirectAsync}.
Currently, GPUDirect Async is only supported on ConnectX NICs.
Additionally, GPUDirect RDMA enables low-latency host-device memcpy via the GDRCopy library~\cite{GDRCopy}.

\subsection{Related Work}

\textbf{Disaggregated Inference\quad}
Splitwise, DistServe, and Mooncake are among the earliest works demonstrating the benefits of disaggregated inference architectures by separating the prefill and decode stages of LLM inference to distinct devices.~\cite{splitwise,DistServe,Mooncake2025}
Our KvCache transfer use case implements this paradigm by bridging \texttt{fabric-lib} to an inference engine to connect prefill and decode clusters via RDMA.

\textbf{Point-to-Point Network Libraries\quad}
\textit{NVSHMEM} exposes both collective operations as well as flexible point-to-point communication.~\cite{NVSHMEM}
It supports both GPU-initiated (IBGDA) and host-proxy communication (IBRC).
However, it suffers from severe performance degradation on EFA.
NVIDIA Inference Xfer Library (NIXL) targets P2P communication for LLM inference, built on UCX~\cite{NIXL,UCX}.
Our production-deployed EFA implementation predates the preliminary EFA support in \textit{NIXL} (v0.6.1, October 2025).
Mooncake also provides an RDMA Transfer Engine, but without support for EFA.
Other libraries, such as \textit{UCCL}~\cite{UCCL} and \textit{MSCCL++}~\cite{MSCCLpp}, focus on network-layer optimizations for collectives rather than point-to-point.

\textbf{Distributed KvCache Storage\quad}
Mooncake Store and DeepSeek 3FS~\cite{3fs} provide distributed storage for KvCaches, albeit they currently lack EFA support.
We complement these systems by providing portable RDMA primitives suitable for cloud deployments.

\textbf{Compute-Communication Overlapping\quad}
Research on overlapping compute with collective communication for LLM kernels~\cite{flux, comet, Triton-distributed, Tilelink} is orthogonal to our focus on RDMA primitives, although we do enable background transfers.

\textbf{LLM Frameworks\quad}
\texttt{fabric-lib} can be integrated into LLM inference frameworks and kernel libraries, such as vLLM, SGLang, TensorRT-LLM, FlashInfer.~\cite{PagedAttention,SGLang,TensorRT-LLM,FlashInfer}
Our P2P weight update approach can be adopted by reinforcement learning frameworks, such as Slime, OpenRLHF, AReaL, veRL, LlamaRL, NVIDIA Nemo.~\cite{LlamaRL, slime_github, mei2025real,fu2025areal,HybridFlow,OpenRLHF}

\section{TransferEngine}

The \textit{TransferEngine} is the core component of \texttt{fabric-lib}, enabling efficient RDMA-based point-to-point communication by abstracting heterogeneous hardware under a simple protocol.
It exposes \textsc{Send}/\textsc{Recv} operations to implement RPC-like interfaces.
For KvCache transfers, it provides paged \textsc{Write}s for bulk writes.
For RL weight transfers, it exposes low-latency high-throughput \textsc{Write} operations.
For MoE routing, it specializes \textsc{Write}s targeting many peers to implement low-latency \textsc{Scatter} and \textsc{Barrier} operations.
There are no ordering guarantees across any of the operations.
An immediate value can be optionally associated with \textsc{Write}s to increment a counter on the receiver upon receipt.

The key contributions of the \textit{TransferEngine} are the portable RDMA interface built on reliable-but-unordered transport semantics and the \textsc{ImmCounter} primitive for order-agnostic completion notification.
The subsections below present the design, API, and engineering optimizations that realize these abstractions.

\subsection{Overview and Design Goals}

The \textit{TransferEngine} exposes a minimal API that abstracts away the complexity of the underlying RDMA interfaces.
It supports multiple interfaces, including EFA with its SRD protocol and a multitude of NICs programmable via \texttt{libibverbs}, including ConnectX-7.
Topologies are transparently detected to handle multiple NICs per GPU: while a single ConnectX-7 NIC provides 400 Gbps bandwidth, achieving equivalent bandwidth on AWS \verb|p5| instances requires aggregating four 100 Gbps EFA NICs (or two 200 Gbps EFA NICs for \verb|p5en| instances).
A single instance spans multiple threads to manage all GPUs and NICs within a node.
Low-latency operations are achieved through zero-copy interfaces and hardware-specific optimizations.

To bridge the gap between the in-order guarantees of RDMA RC and the out-of-order delivery of EFA SRD, the \textit{TransferEngine} relies solely on a novel \textsc{ImmCounter}.
Instead of relying on ordering guarantees, all completion notifications are handled by polling completion queues and delivering notifications upon the bulk receipt of immediate values. 
Notifications are delivered through callbacks or atomic flags.

\subsection{Architecture}

\begin{figure}[t]
\centering
\includegraphics[width=\linewidth]{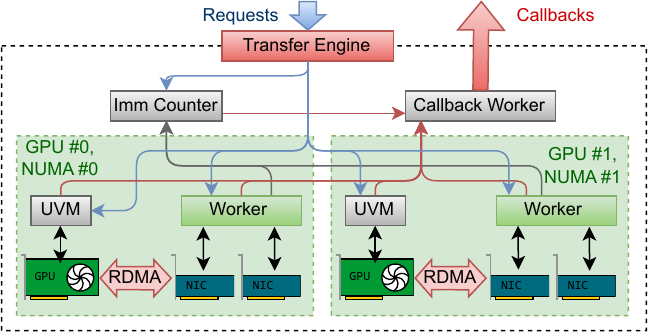}
\caption{\textit{TransferEngine} managing GPUs across NUMA nodes, each with multiple NICs. Commands are forwarded to workers, which respond back to the callback handler or \textsc{ImmCounter}.}
\label{fig:transfer-engine}
\end{figure}
\begin{figure}[t]

\begin{minted}[baselinestretch=0.75,fontsize=\fontsize{7pt}{7pt}\selectfont]{rust}
#[serde] struct NetAddr(Bytes);
#[serde] struct MrDesc{ ptr: u64, rkeys: Vec<(NetAddr, u64)> }
struct MrHandle(NonNull<c_void>); type Offset = u64;
struct Pages{ indices: Vec<u32>, stride: u64, offset: Offset }
struct PeerGroupHandle(u64);
struct ScatterDst{ len: u64, src: Offset, dst: (MrDesc,Offset)}
enum OnDone { Callback(fn () -> ()), Flag(Atomic<bool>) }

trait TransferEngine {
  fn main_address() -> NetAddr;
  // Memory Region Management
  fn reg_mr(ptr, len, device) -> (MrHandle, MrDesc);
  // Two-sided Send/Recv
  fn submit_send(addr: NetAddr, msg: &[u8], cb: fn () -> ());
  fn submit_recvs(len: u64, cnt: u64, cb: fn (&[u8]) -> ());
  // One-sided Write
  fn expect_imm_count(imm: u32, count: u32, cb: fn () -> ());
  fn submit_single_write(len: u64, imm: Option<u32>
      src: (MrHandle, Offset), dst: (MrDesc, Offset), OnDone);
  fn submit_paged_writes(page_len: u64, imm: Option<u32>,
      src: (MrHandle, Pages), dst: (MrDesc, Pages), OnDone);
  // One-sided Write to a group of peers
  fn add_peer_group(addrs: Vec<NetAddr>) -> PeerGroupHandle;
  fn submit_scatter(h: Option<PeerGroupHandle>, OnDone,
      imm: Option<u32>, src: MrHandle, dst: Vec<ScatterDst>);
  fn submit_barrier(h: Option<PeerGroupHandle>, OnDone,
      imm: u32, dst: Vec<MrDesc>);
  // Watcher for CPU-GPU synchronization
  fn alloc_uvm_watcher(cb: fn(u64,u64) -> ()) -> NonNull<u64>;
}
\end{minted}
\caption{\textit{TransferEngine} API pseudo-code. Error handling, domain sharding, and resource releasing are omitted.}
\label{code:api}
\end{figure}

\Cref{fig:transfer-engine} illustrates the architecture of the \textit{TransferEngine}.
The engine spawns a worker per GPU, with each managing a generic \textsc{DomainGroup} that coordinates all the associated RDMA NICs, typically 1--4 NICs depending on the hardware platform.
Within a generic \textsc{DomainGroup}, each \textsc{Domain} is specialized to the hardware and responsible for a single NIC, handling queue pair management, work submission, and completion polling.

Each \textit{TransferEngine} instance exposes a single \emph{main address} for identification and discovery to remote peers.
We use a \verb|NetAddr| struct to capture and serialize network address of a \textsc{Domain}, exchanging the structure between peers that wish to communicate.
As a restriction, all peers must use the same number of NICs per GPU.
Consequently, any transfer has full knowledge of the NICs between the source and destination domain, allowing the \textit{TransferEngine} to shard or balance the request.
This is crucial for EFA, which achieves the full 400Gbps bandwidth using multiple adapters.

\subsection{API Design}

The API exposed by the \textit{TransferEngine} is outlined in \Cref{code:api}, implementing the abstractions over RDMA:

\textbf{Memory Registration\quad} Memory regions must be registered with the engine, returning a serializable \verb|MrDesc| that can be exchanged with peers to submit \textsc{Write}s through and a local \verb|MrHandle| to be used as the source of transfers.
The same API can register both host-side buffers and GPU memory.
Behind the opaque types, the handles carry the addresses of all the NICs they are associated with, along with the domain-specific remote keys attached to them as a list of (\verb|NetAddr|, \textsc{RKey}) pairs.

\textbf{Point-to-Point Transfer\quad} \texttt{submit\_send} and its remote counterpart \texttt{submit\_recvs} wrap the \textsc{Send}/\textsc{Recv} operations to expose RPC-style communication, exchanging small payloads.
Submission involves a copy to allow the caller to reuse or release the buffer immediately.
Receive posts a rotating pool of buffers to receive data into.
Upon each message, a buffer is temporarily taken out of the pool to allow the callback to process it without copying.
Upon callback completion, it is automatically re-posted.
Sufficient buffers must be allocated to avoid rejecting messages.
These operations utilize only the first NIC in a domain group.

\texttt{submit\_single\_write} and \texttt{submit\_paged\_writes} transfer data from a source to a destination buffer, writing contiguous or paged regions determined by indirect indices, strides and offsets, respectively.
The engine translates these operations into possibly multiple zero-copy one-sided RDMA \textsc{Write} operations, sharding and rotating them along the available NICs.
Each Transfer can optionally carry a 32-bit immediate number to notify the receiver upon completion.
Transfer completion notifications are delivered asynchronously to the caller.

\texttt{submit\_scatter} send a slice from a source buffer to each peer in a peer group at different offsets in their receive buffers, while \texttt{submit\_barrier} is an immediate-only operation for peer notification.
These are optimized wrappers around \textsc{Write}, allowing applications to pre-register a \emph{PeerGroup} to target with low-latency bulk transfers.

Transfers execute between two devices: it is up to the user to coordinate operations across multiple devices in a system.

\textbf{UVM Watcher\quad} In order for the host to be able to initiate transfers upon GPU progress, \texttt{alloc\_uvm\_watcher} registers a callback invoked upon changes to a word in memory.
It allocates a unified virtual memory (UVM) location that can be updated by the devices, including kernels within a CUDA graph, which is continuously polled by a CPU thread using GDRCopy.
Since not all changes are guaranteed to be observed immediately, the callback is invoked the old and the new values, allowing it to respond to GPU-side progress.

\textbf{Completion Notification\quad} Notifications are delivered for both send confirmation and receive completion, either through callbacks or atomic flags.
The \textsc{ImmCounter} is a dedicated component that keeps track of per-immediate counters incremented on events retrieved from the completion queue of the underlying devices.
Events are generated on the sender after a transfer is completed and on the receiver once a payload with an immediate attached to it has been fully delivered, guaranteeing atomicity.
The counters are allocated in the same NUMA node as the domain worker.

The counters can either be transparently synchronized with the GPU via GDRCopy, observed directly through polling or handed off to a callback on a separate dedicated thread within the engine, registered using \texttt{expect\_imm\_count}.
All synchronization is implemented using such counters, as otherwise there are no other ordering guarantees.

Correctness of \textsc{ImmCounter} under unordered transport relies on PCIe ordering guarantees:
the RDMA specification requires that the data payload of a \textsc{WriteImm} is issued before the immediate value, and the PCIe switch guarantees ordering of writes targeting the same device.
Although the immediate value targets the CPU while the data targets the GPU, the host-proxy architecture avoids data races: after the CPU observes the target \textsc{ImmCount}, any subsequent CPU-to-GPU transaction (e.g., launching a kernel or updating a flag via GDRCopy) is ordered by the PCIe switch after the preceding NIC-to-GPU data writes, ensuring the payload is visible to the GPU.

\subsection{Implementation}

Our \textit{TransferEngine}, implemented in Rust, carefully optimizes allocations, threading and synchronization to minimize latency in order to achieve high throughput.
The engine spawns one worker thread per \textsc{DomainGroup},
each pinned to a CPU core on the NUMA node to which the devices are attached to minimize both scheduling and memory access latency.
Data structures specific to a domain are allocated after pinning to ensure that memory is reserved in the correct NUMA node.
One worker thread is responsible for handling up to 4 \textsc{Domain}s, each managing a single NIC, whereas another dedicated thread is responsible for polling the GPU to update UVM watchers.
Cross-thread communication is done through lock-free queues.

The API forwards requests to the \textsc{DomainGroup} serving the appropriate device, with the first one also serving the host.
In a tight loop, the domain worker polls for new work, prioritizing the submission of new requests.
Depending on the hardware and configuration, requests are sharded and load-balanced across the available \textsc{Domain}s.
The first \textsc{Write} of a composite request is immediately posted to the NIC's send queue in the \textsc{Domain}.
Once new requests are exhausted, the worker proceeds to progress on pending, posting writes to fill up the hardware pipeline.
Subsequently, completion queues are polled to query for finished transfers and immediate counter increments.
Events are aggregated to deliver per-transfer notifications, handing the transfer over to a dedicated callback thread shared by all groups.

Sharding inside a \textsc{DomainGroup} is flexible.
Transfers can target specific NICs by index.
A single \textsc{Write} can be split.
Paged transfers, scatter and barrier operations, which all translate to multiple \textsc{Write}s, can shard across all NICs.

\subsection{Hardware-Specific Optimizations}

The \textsc{Domain}s within the \textit{TransferEngine} are specialized and optimized for the hardware under their control:

\textbf{AWS EFA\quad} We implement EFA support using \texttt{libfabric}, managing a fabric domain per NIC within the \textsc{DomainGroup}.
Since EFA diverges from the RDMA spec which does not require a valid target descriptor for immediate-only zero-sized writes, we enforce a valid descriptors for all transfers.
For bulk transfers and peer groups, we employ work request (WR) templating, pre-populating and retaining the common fields of \texttt{libfabric} descriptors before posting.

\textbf{NVIDIA ConnectX-7\quad} We implement ConnectX support through \texttt{libibverbs}.
For each peer, we use an UD queue pair to exchange RC hanshakes.
We create 2 RC queue pairs per peer: one for two-sided \textsc{Send}/\textsc{Recv} operations and another for one-sided \textsc{Write} and \textsc{WriteImm} operations.
This separation is necessary because both \textsc{Recv} and \textsc{WriteImm} completions consume work requests
in posting order.
This allows us to provide high-level \textsc{Recv} semantics while supporting \textsc{WriteImm} without interference.
In addition to WR templating, we employ WR chaining by linking up to 4 work requests through the \verb|next| pointer of \texttt{ibv\_send\_wr}, reducing the number of doorbell rings to the NIC.
Additionally, we enable \texttt{IBV\_\-ACCESS\_\-RELAXED\_\-ORDERING} to permit out-of-order PCIe transactions between the NIC and GPU memory, reducing latency.

\section{KvCache Transfer}
\label{sec:kvcache}

In this section, we outline a production-tested implementation of disaggregated inference relying on the \textit{TransferEngine}.
Pseudocode is provided in \Cref{sec:appendix-kvcache}.
In disaggregated mode, a prefiller node runs prefill on the input tokens, transferring the resulting KV pages and any additional context, such as last token hidden states and logits for speculative decoding, to the decoder node which proceeds to decode tokens one-by-one.

\begin{figure}[t]
\centering
\includegraphics[width=\linewidth]{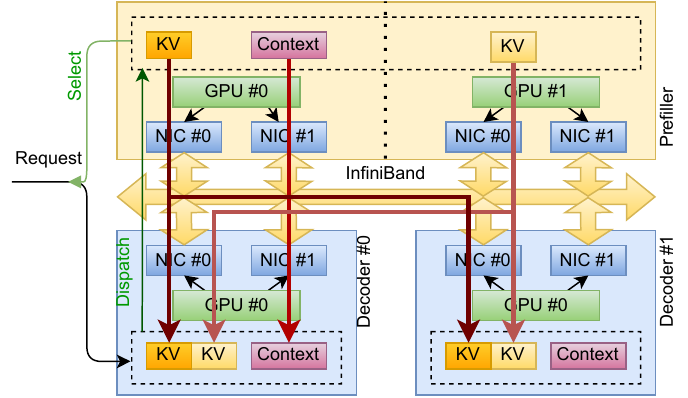}
\caption{KV transfer between prefillers and decoders}
\label{fig:kv_transfer}
\end{figure}

Figure~\ref{fig:kv_transfer} illustrates our disaggregated setup.
Once a request is received, a global scheduler selects a prefiller node and a decoder node to process it and forwards the request to the decoder.
The decoder pre-allocates KV pages and storage for any context before
dispatching the request to the designated prefiller using \texttt{submit\_send}, indicating the KV page indices where contents should be transferred to.

During chunked prefill, we increment the UVM watcher value after the attention output projection of each layer, which is CUDA Graph compatible.
Once \textit{TransferEngine} detects the change in the UVM watcher value, the layer's transfer is initiated, copying over the pages from the current chunk via \texttt{submit\_\-paged\_\-writes}.
When the last chunk is complete and the context has been populated, it is copied over via \texttt{submit\_\-single\_\-write}.
Prefillers await for commands using \texttt{submit\_\-recvs}.

KV transfers must account for differences in the sharding or replication of KvCaches between prefillers and decoders. 
MLA \cite{liu2024deepseek} replicates the compressed KvCache entries if tensor parallelism is used: under such a scheme, prefiller ranks are randomly matched with decoder ranks to balance the transfers of replicas.
Under GQA \cite{ainslie2023gqa}, we rely on page-wise offsets and strides to select slices from the source KvCache to copy into corresponding offsets in the destination KvCache.
To minimize the number of writes and ensure that individual writes are sufficiently large, the KvCaches are laid out with heads preceding the pages, ensuring continuity within consecutive heads.

The prefiller does not issue an explicit completion message: the decoder knows in advance the number of transfers it expects and uses \texttt{expect\_\-imm\_\-count} to be notified by the \textit{TransferEngine} of transfer completion and start decoding.

The complexity of a production-ready implementation of disaggregated decoding lies in the handling of errors and cancellation.
Cancellation triggerred by a decoder must be explicitly confirmed by the prefiller, as the KV pages cannot be reused as long as there is a possibility of a remote write clobbering them.
We rely on heartbeat messages between prefillers and decoders to detect transport layer failures.
If a prefiller node is unresponsive, requests are cancelled on the decoder after a timeout, as transfers can no longer reach it.
A per-request cancellation token can stop all future transfers whilst also waiting after all pending operations before sending the cancellation confirmation.

\section{RL Rollout Weight Transfer}
\label{sec:rollout}

\begin{figure}[t]
\centering
\includegraphics[width=\linewidth]{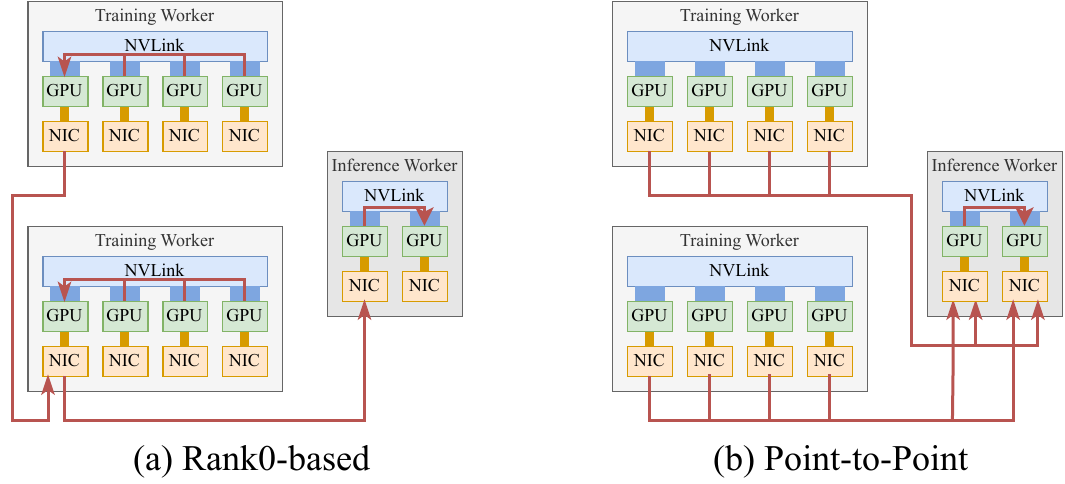}
\caption{Weight transfer data path for different approaches.}
\label{fig:rl-rank0-vs-p2p}
\end{figure}

\begin{figure}[t]
\centering
\includegraphics[width=\linewidth]{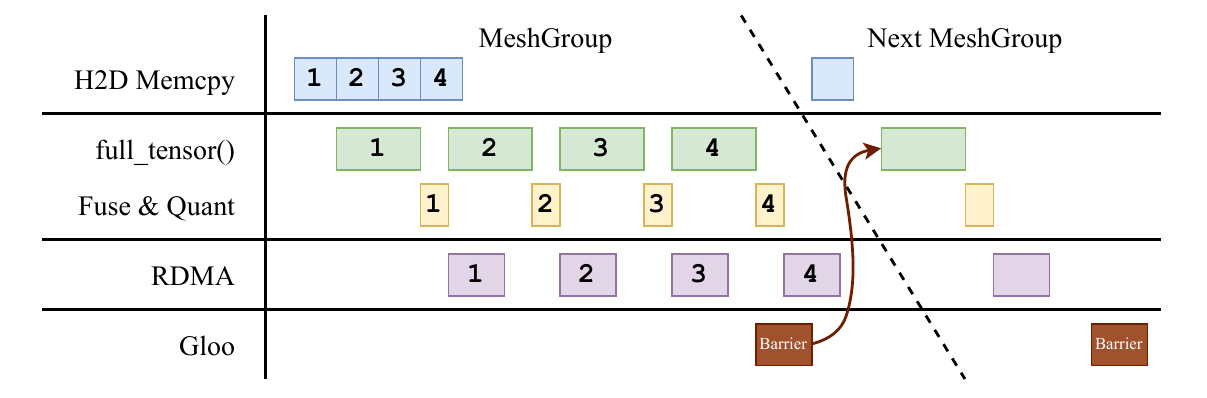}
\caption{Pipelined weight transfer execution.}
\label{fig:rl-weight-transfer-pipeline}
\end{figure}

In asynchronous reinforcement learning fine-tuning, training and inference run on separate GPUs.
After each training step, new weights must be pushed to inference nodes, which can take tens to hundreds of seconds on trillion-parameter models using existing frameworks.
Our solution achieves 1.3-second cross-machine parameter updates for models at the scale of Kimi-K2 (1T parameters), DeepSeek V3 (671B) and Qwen3 (235B)~\cite{Kimi-K2,DeepSeek-V3,Qwen3}, transferring weights from 256 training GPUs (\texttt{bf16}) to 128 inference GPUs (\texttt{fp8}).

\subsection{Point-to-Point Weight Transfer}

Existing frameworks tend to form a global collective communication world for all training and inference GPUs.
Weights are gathered to training sub-group \textit{Rank0}, then broadcast to \textit{Rank0} of each inference sub-group, bottlenecked by the NIC of training \textit{Rank0}. In contrast, in our P2P approach each training GPU send weights directly to inference GPUs via one-sided RDMA \textsc{Write}, utilizing the full cluster bandwidth across all NICs.
\Cref{fig:rl-rank0-vs-p2p} illustrates the difference between the two approaches.

At initialization, the controller script performs three steps:
First, it gathers parameter metadata from all training and inference GPUs, including weight name, shape, dtype, and DTensor sharding.
Next, it computes a static weight transfer schedule, mapping which training GPU sends which parameter to which inference GPU, and in what order.
Finally, it broadcasts the schedule to all training GPUs.
\Cref{sec:appendix-rl} provides pseudocode for this workflow.

At each training step, the controller signals training GPUs to begin sending weights.
The inference nodes remain unaware of the transfer, as it uses one-sided operations.

\subsection{Pipelined Weight Transfer Execution}

Our training job shards model weights using FSDP~\cite{FSDP}.
Different parameter types (e.g., MoE vs non-MoE) require different FSDP sharding strategies.
Each sharding strategy partitions the global DeviceMesh into disjoint sub-meshes.
We call each sub-mesh a \emph{MeshGroup}.
Parameters within a MeshGroup are transferred in parallel, while MeshGroups are processed sequentially.

We treat the transfer of each parameter tensor as a \textit{task}.
To utilize different hardware resources simultaneously, we split each task into four pipeline stages that overlap in time, as illustrated in \Cref{fig:rl-weight-transfer-pipeline}:
(1) Host-to-device memcpy if FSDP offloads weight to CPU.
(2) Parameter preparation: Reconstruct full weight with \texttt{full\_tensor()}, apply projection fusion, quantize if needed.
(3) RDMA transfer: Zero-copy \textsc{Write} to remote inference GPU memory.
(4) Global barrier: After all \texttt{full\_tensor()} calls are done, synchronize across mesh groups using GLOO via Ethernet.

\texttt{full\_tensor()} and other GPU operations introduce extra GPU memory usage.
To avoid out-of-memory errors, we only start a new task if the current in-flight tasks occupy less temporary GPU memory than a configurable watermark.

\section{MoE Dispatch/Combine}
\label{sec:moe}

\begin{figure*}[t]
\includegraphics[width=\linewidth]{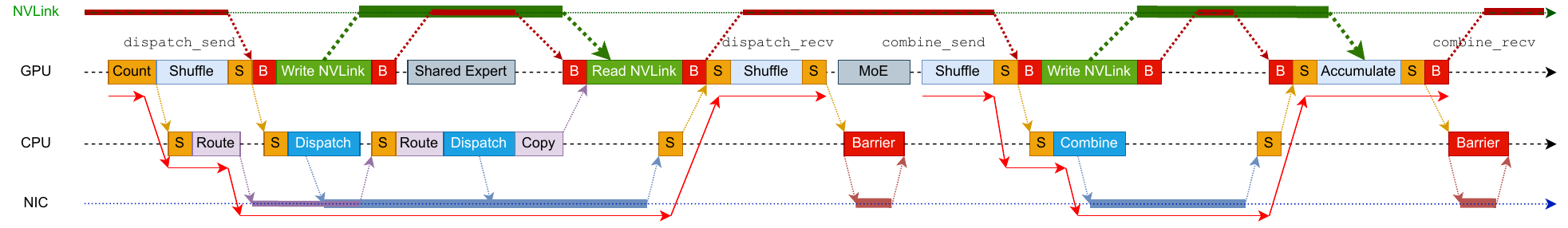}
\caption{Dispatch and Combine GPU-CPU-NIC coordination}
\label{fig:timing}
\end{figure*}

We present a set of low-latency kernels for MoE dispatch and combine built around the \textit{TransferEngine}, relying on a host proxy thread to coordinate the GPUs and the NICs.
Within a node, we also utilize NVLink to reduce the load on the network.
Despite the added latency of a proxy thread, we achieve state-of-the-art decode performance while maintaining competitive performance for prefill without any tweaks.

These kernels demonstrate the feasibility of proxy-based MoE dispatch with support for a wider range of network cards, such as EFA.
Consequently, we focus on decode performance (up to 128 tokens per rank) as it is latency-bound, suffering more significantly from the added PCIe, driver and firmware overheads across the devices involved.

\subsection{Architecture}

Routing is implemented using split kernels for both decode and combine, with a sender half preparing the data into send buffers to be written to peers and a receiver half shuffling tokens from receiver buffers into tensors used by other kernels.
The individual kernels fully utilize all SMs and the memory bandwidth of the GPU to reduce latency, however their runtime is short enough to interleave other work between them, enabling overlapping and micro-batching.
The host proxy uses GDRCopy to poll the GPU for progress, invoking the \textit{TransferEngine} when source buffers are ready.

Our design, illustrated in Figure~\ref{fig:timing}, minimizes the proxy overhead by reducing the number of writes issued.
To reduce the GPU memory required for send and receive buffers, all peers first exchange routing information, per-expert token counts, to determine a unique range in a contiguous receive buffer to write to.
Since these payloads are small, their latency is hidden by speculatively dispatching a small number of tokens into private per-source buffers.
Combine issues a single scatter as it re-uses the routing information.
Routing information is always handled by the \textit{TransferEngine}, but payloads can be copied via NVLink within the same node. 
Consequently, each rank issues up to 2 \textsc{Write}s for dispatch and 1 \textsc{Write} for combine for each inter-node peer.

Receiver buffers must be sized to account for all tokens being sent to the current rank.
Assuming there are $N$ ranks hosting $E$ experts, each dispatching $T$ tokens to $R$ experts, the upper bound is $N \cdot T \cdot max(R, \frac{E}{N})$.
Senders pack writes into such a contiguous buffer instead of relying on larger per-rank receiver buffers.
To minimize overhead, the larger dispatch receive buffer can be re-used by combine send.

\subsection{Dispatch}

\begin{figure}[t]
\centering
\includegraphics[width=\linewidth]{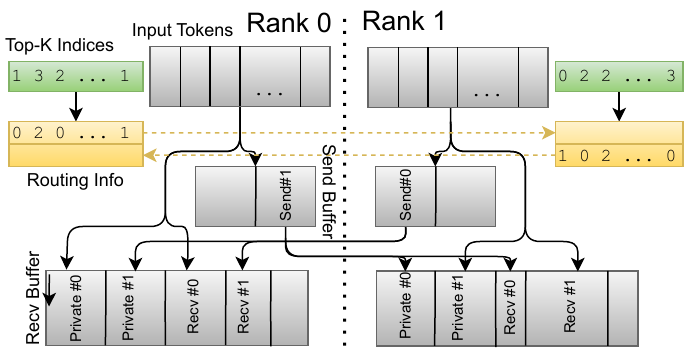}
\caption{Dispatch into private and contiguous buffers}
\label{fig:buffers}
\end{figure}

The dispatch kernel receives the tokens and the $R$ indices of the experts to which they must each be routed, returning a tensor which packs the tokens from all the ranks assigned to the local experts.
The kernel first counts the number of tokens sent to each expert in shared memory and transfers the counts to the host via unified memory.
It then signals the proxy, which uses the \textit{TransferEngine} to initiate the scatter of the routes to all peer ranks.
The input tokens are then copied into the send buffers, creating a contiguous source to be scattered to each peer, as shown in Figure~\ref{fig:buffers}.

The proxy is notified to scatter tokens up to a fixed limit into private buffers on each receiver peer.
For decode-sized batches, the latency from the launch of the dispatch kernel to the first transfer is around $\SI{15}{\micro\second}$ assuming EP=64.
At this point, the transfers of routes and tokens saturate the NIC bandwidth. 
Once the routes are received and the position of tokens from each source rank can be determined on each destination rank, the remainder of the tokens are scattered to the peers, placing them contigously into a shared receiver buffer.
The host-side work to process routes and dispatch the second round of transfers takes tens of microseconds.
The number of tokens in the initial transfer is chosen to hide this overhead.
Although this latency is not on the critical path, reducing it can further reduce the size of the private buffers.
Once all transfers are acknowledged as having completed and all incoming writes are accounted for, a barrier synchronizes the proxies via the \textit{TransferEngine}.

Following the dispatch of transfers to inter-node peers over RDMA, the send kernel proceeds to issue writes over NVLink to transfer tokens within the same node, while RDMA transfers are pending in the background.
NVLink peers synchronize themselves through their own set of barriers: before writing to a peer, each rank must ensure that the data has been read and can be overwritten.
This is done through flags written and read with \texttt{relaxed} semantics.
After the payloads are written, ranks synchronize through \texttt{release}-\texttt{acquire} flags with each other.

Write ordering in send kernels is latency-critical due to the lack of granularity in memory barriers.
NVLink is exposed via virtual memory, transparently translating reads and writes to peer devices mapped into the current address space into transactions on the interconnect.
Loads are universally expensive, as they block the execution pipeline until they are satisfied.
In contrast, stores are fire-and-forget, until a memory barrier is encountered, which blocks until all prior stores within their scope complete.
Since both the host system and NVLink peers are within the same scope,
a barrier ensuring ordering with the host might be slowed down by previously issued writes over NVLink.
This is avoided by first signalling the host, then issuing NVLink writes after a grid barrier.
This strategy increases the total execution time of send kernels, but it reduces latency on the critical path to the first RDMA transfer.

With NVLink, it is usually preferable to push data from a source device to a target, saving a trip time.
Additionally, after the stores are acknowledged on the current device, useful work can be executed while the transfers are in-flight to the remote.
With dispatch, we only push the tokens to the private receive buffers, since at this stage the centralized routing information is not available to determine exactly where the rest of the tokens should be placed on the peer.
The receiver half of the kernel kicks off by synchronizing on the barrier and reading the remaining tokens.
These loads are likely to complete before the RDMA operations.

After dispatch send, transfers run in the background. 
Once the receiver kernel is called, it waits for the \textit{TransferEngine} to report the completion of all transfers via \textsc{ImmCounter} and GDRCopy.
Relying on indices computed based on the routing information, tokens are
re-ordered with optional padding between experts following a layout suitable for Grouped GEMM kernels.
Since the receiver kernels must process around $T \cdot R$ tokens while on the critical path, work is split across all available SMs and pipelining is maximised to fully utilize the available HBM bandwidth.

\subsection{Combine}

Since routing information is centralized during the dispatch stage, combine
transfers all payloads in a single scatter.
The command is prepared during the idle time between send and combine, during which the GPU would typically execute a grouped GEMM.
The sender, similarly to dispatchs, prepares send buffers and pushes tokens via NVLink to intra-node peers.
The host proxy is signalled to send the \textsc{Scatter} requests to the \textit{TransferEngine}, after which the kernel finishes execution.
The receiver caches all relevant offsets derived from routing information in shared memory before waiting for all tokens to be received.
The weighted average of the tokens is then computed locally on each rank.

The combine stage re-uses the same buffers as the dispatch stage, thus it requires both an NVLink and an RDMA barrier ensuring the completion of all prior operations before overwriting the send buffers.
At this point, the host proxy also waits for the \textit{TransferEngine} to confirm that all writes have been sent out.
Particularly for EFA, which also waits for receipt confirmation, it is important to maximise the interval between posting a write and checking its status.

\subsection{Comparison with DeepEP}

DeepEP offers state-of-the-art latency, however they are tied to ConnectX due to their reliance on IBGDA and \verb|mlx5| driver.
Our kernels are more portable, relying on a host proxy, supporting both EFA and ConnectX.
Despite the added overhead, our latency exceeds DeepEP.

DeepEP kernels rely on the strong ordering guarantees of RC QP.
The tokens are balanced across the available SMs which transfer them over a QP one-by-one.
This ensures a lower latency to the first transfer, however it also involves more work on a per-token basis and results in more packets being sent over the network. 
Token counts and completion are signalled via \textsc{Atomic}s.
In contrast, our kernels spend more time before the first transfer is initiated due to the additional GPU-CPU-NIC communication overheads over PCIe. However, bulk transfers achieve better network utilization.

For prefill, we scale our single-transfer strategy without tweaks.
In contrast, DeepEP achieves better latencies by pre-accumulating tokens via NVLink on the sender node, reducing the amount of data transferred.
Additionally, the DeepEP kernels use less buffer memory as subsets of tokens are transferred in batches.
While this approach is faster, it also has implications on accuracy and determinism, as accumulation it not done entirely on \texttt{fp32} in a fixed order.

\section{Evaluation}

We evaluate on two cluster configurations: 8$\times$H200 nodes with 2$\times$\SI{200}{Gbps} EFA per GPU, and 8$\times$H100 nodes with \SI{400}{Gbps} ConnectX-7 per GPU.
End-to-end evaluations run on a custom inference engine built on PyTorch.

\subsection{Point-to-Point Communication}

\begin{figure}[ht]
\centering
\includegraphics[width=\linewidth]{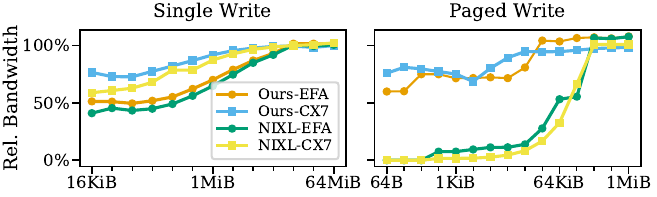}
\caption{Point-to-Point communication performance}
\label{fig:p2p-bw}
\end{figure}

\begin{table}[ht]
\small
\begin{center}
\resizebox{\linewidth}{!}{%
\begingroup
\setlength{\tabcolsep}{2pt}
\begin{tabular}{l|r|rr|rr}
\multicolumn{2}{c}{} & \textbf{EFA} & \multicolumn{1}{c}{} & \textbf{CX-7} &  \\
\hline
\hline
\multirow{4}{*}{\rotatebox[origin=c]{90}{\textbf{Single}}}
& 64 KiB     & 16 Gbps  &            & 44 Gbps  &            \\
& 256 KiB    & 54 Gbps  &            & 116 Gbps &            \\
& 1 MiB      & 145 Gbps &            & 245 Gbps &            \\
& 32 MiB     & 336 Gbps &            & 378 Gbps &            \\
\hline
\multirow{4}{*}{\rotatebox[origin=c]{90}{\textbf{Paged}}}
& 1 KiB      &  17 Gbps & 2.11M op/s &  91 Gbps & 11.10M op/s \\
& 8 KiB      & 138 Gbps & 2.10M op/s & 320 Gbps &  4.89M op/s \\
& 16 KiB     & 274 Gbps & 2.08M op/s & 367 Gbps &  2.80M op/s \\
& 64 KiB     & 364 Gbps & 0.69M op/s & 370 Gbps &  0.71M op/s \\
\hline
\end{tabular}
\endgroup
}
\end{center}
\caption{EFA and ConnectX-7 performance comparison}
\label{table:p2p-efa-cx7}
\end{table}

The performance of KvCache transfers and RL rollouts is determined by the throughput of single and paged writes.
We compare higher-level libraries (\textit{TransferEngine} and \textit{NIXL} v0.6.1) against hardware benchmarks for the underlying NICs, measuring the fraction of bandwidth for a range of message sizes.
On ConnectX we compare against \texttt{ib\_\-write\_\-bw} from \textit{rdma-core}.
On EFA, we use \texttt{fi\_\-rma\_\-bw} from \textit{libfabric} to measure peak single-NIC bandwidth, which we double to match the engine running on two NICs.

\Cref{fig:p2p-bw} shows the fraction of the peak bandwidth achieved by both libraries.
To saturate with single \textsc{Write}, messages of at least 16MiB are required, while 32KiB and 64KiB are sufficient to saturate with paged \textsc{Write} on both \textit{TransferEngine} and \textit{NIXL}, respectively.
Performance between our solution and \textit{NIXL} is relatively close, with the \textit{TransferEngine} being slightly faster.
EFA requires larger messages to saturate, explaining the gap observed on MoE routing.

\Cref{table:p2p-efa-cx7} details absolute performance numbers.
For 256 KiB single \textsc{Write} (typical for our MoE routing), our \textit{TransferEngine} achieves 54 Gbps on EFA and 116 Gbps on ConnectX-7.
At 64 KiB paged \textsc{Write} (typical size for a KvCache page), both EFA and ConnectX-7 are able to saturate the available bandwidth.
The size of RL weight transfer messages is well beyond the saturation point.

\subsection{KvCache Transfer}

\begin{table}[ht]
\small
\centering
\setlength{\tabcolsep}{4pt}
\begin{tabular}{r rr rr rr}
\toprule
 & \multicolumn{2}{c}{TTFT (\SI{}{\ms})} & \multicolumn{2}{c}{Per-layer (\SI{}{\ms})} & & \\
\cmidrule(lr){2-3} \cmidrule(lr){4-5}
Seqlen & Non- & Disagg & Compute & Transfer & Steps & Pages \\
\midrule
4K   & $214$   & $260$   & $2.267$  & $0.661$ & $1$ & $256$ \\
8K   & $433$   & $501$   & $4.578$  & $0.952$ & $1$ & $512$ \\
16K  & $929$   & $1042$  & $9.860$  & $1.610$ & $1$ & $1024$ \\
32K  & $2179$  & $2317$  & $13.295$ & $1.606$ & $2$ & $1024$ \\
64K  & $5681$  & $5852$  & $20.344$ & $1.611$ & $4$ & $1024$ \\
128K & $16735$ & $17056$ & $34.895$ & $1.609$ & $8$ & $1024$ \\
\bottomrule
\end{tabular}
\caption{KvCache transfer impact on TTFT (Qwen3-235B, H200 TP4, 2$\times$\SI{200}{Gbps} EFA)}
\label{table:kvcache-ttft}
\end{table}

\Cref{table:kvcache-ttft} shows the end-to-end impact of disaggregated KvCache transfer on TTFT for Qwen3-235B on H200 with TP4 and 2$\times$\SI{200}{Gbps} EFA per GPU.
We use a KvCache page size of \SI{32}{\kilo\byte} (128 tokens) with chunk-prefill length up to 16384 tokens, CUDA Graph enabled, and UvmWatcher to detect layer completion.
Although 1024 pages of \SI{32}{\kilo\byte} cannot saturate RDMA bandwidth, layer-by-layer KvCache transfer is still hidden by computation.
The observed TTFT slowdown relative to non-disaggregated execution is mainly from our inference engine performing one extra decode pass for the final input token, rather than from KvCache transfer.

\begin{table}[ht]
\small
\centering
\setlength{\tabcolsep}{3.5pt}
\begin{tabular}{l rrrrrrr}
\toprule
Callback & avg & min & p50 & p90 & p99 & p99.9 & max \\
\midrule
Rust   & $6.3 \pm 1.3$ & $2.5$ & $6.2$ & $7.0$  & $12.6$ & $19.4$  & $64.8$ \\
Python & $9.8 \pm 9.0$ & $6.1$ & $9.3$ & $11.1$ & $20.4$ & $41.7$ & $3325.0$ \\
\bottomrule
\end{tabular}
\caption{UvmWatcher callback latency under CUDA Graph (\SI{}{\micro\second})}
\label{table:uvmwatcher}
\end{table}

\Cref{table:uvmwatcher} reports the tail latency of UvmWatcher callbacks under CUDA Graph.
When the callback is in Rust, latency is tightly bounded, only slightly above the \SIrange{2}{5}{\micro\second} PCIe latency.
With Python, overhead is still acceptable.

\subsection{RL Weight Transfer}

\begin{table}[ht]
\small
\centering
\setlength{\tabcolsep}{5pt}
\begin{tabular}{l rrr}
\toprule
Operation & Time & Avg. per-call & Count \\
\midrule
Total & \SI{1233}{\ms} & --- & --- \\
\quad Memcpy H2D & \SI{184}{\ms} & \SI{378}{\micro\second} & 487 \\
\quad \texttt{full\_tensor()} & \SI{518}{\ms} & \SI{532}{\micro\second} & 974 \\
\quad Fuse projections & \SI{18}{\ms} & \SI{37}{\micro\second} & 487 \\
\quad Quantize & \SI{88}{\ms} & \SI{137}{\micro\second} & 647 \\
\quad RDMA submit & \SI{26}{\ms} & \SI{23}{\micro\second} & 1144 \\
\quad Waiting for other ranks & \SI{357}{\ms} & --- & --- \\
\quad Remaining\textsuperscript{*} & \SI{42}{\ms} & --- & --- \\
\bottomrule
\multicolumn{4}{l}{\textsuperscript{*}\scriptsize Includes extra RDMA time not hidden by other operations.}
\end{tabular}
\caption{RL weight transfer latency breakdown for one rank}
\label{table:rl-breakdown}
\end{table}

\Cref{table:rl-breakdown} shows a per-rank latency breakdown for transferring Kimi-K2-1T weights from 256 training GPUs (BF16, \mbox{FSDP/PP/EP=16/2/8}) to 128 inference GPUs (FP8, EP=32), profiled with PyTorch profiler.
The total transfer completes in \SI{1.2}{\second}, an order of magnitude faster than the \SIrange{10}{100}{\second} reported by prior systems for comparable model sizes~\cite{MoonshotAI-Checkpoint-Engine,NeMo-RL-PR1267,slime-issue132,BiaoHe-Optimizing-Weight-Sync-in-slime}.

The breakdown confirms effectiveness of the pipelining shown in \Cref{fig:rl-weight-transfer-pipeline}.
The critical path is dominated by \texttt{full\_tensor()} for FSDP parameter unsharding (\SI{518}{\ms}) and synchronization across ranks (\SI{357}{\ms}).
RDMA transfers run concurrently with parameter preparation; only \SI{42}{\ms} of transfer time falls outside the pipeline overlap.
The CPU overhead of posting 1144 RDMA work requests is \SI{26}{\ms}.
Our training logs for DeepSeek-V3-671B and Qwen3-235B show similar transfer times of \SIrange{1.2}{2}{\second}.

\subsection{MoE Dispatch/Combine}

We evaluate the performance of the MoE kernels for both decode and prefill across 8, 16, 32 and 64 GPUs, comparing them to DeepEP, which is the most performant open-source implementation, as well as the portable and open-source \textit{pplx-kernels}~\cite{pplx-kernels} which are built around NVSHMEM v3.4.5, running on both EFA and ConnectX-7.
While there is overlap between our work and UCCL-EP~\cite{UCCL-EP}, we do not include a detailed comparison as their published latencies are substantially higher.
We show that on ConnectX-7 adapters our performance is the new state-of-the-art, despite the use of the host proxy.

\subsubsection{End-to-End Decode Speed}

\begin{table}[ht]
\small
\centering
\setlength{\tabcolsep}{5.5pt}
\begin{tabular}{llrrr}
\toprule
Cluster & Kernel & \multicolumn{1}{c}{batch=2} & \multicolumn{1}{c}{batch=8} & \multicolumn{1}{c}{batch=32} \\
\midrule
\multirow{2}{*}{H200 EFA} & Ours & $66.752$ & $56.459$ & $32.003$ \\
 & \textit{pplx-kernels} & $20.972$ & $11.607$ & $4.903$ \\
\midrule
\multirow{2}{*}{H100 CX-7} & Ours & $78.420$ & $67.666$ & $36.066$ \\
 & DeepEP & $73.758$ & $65.785$ & $36.253$ \\
\bottomrule
\end{tabular}
\caption{End-to-end MoE decode speed (tokens/s, DeepSeek-V3, MTP, EP=DP=64)}
\label{table:e2e-moe-decode}
\end{table}

\Cref{table:e2e-moe-decode} reports end-to-end decode speed of DeepSeek-V3 with MTP (draft length 1, acceptance rate 80\%) at EP=DP=64.
On EFA, our kernels achieve 3--6$\times$ higher throughput than \textit{pplx-kernels}, making real-time inference practical on EFA for the first time.
On ConnectX-7, our kernels match or slightly exceed DeepEP across all batch sizes, despite using a host proxy rather than GPU-initiated RDMA.
The two implementations share the same inference engine and differ only in the MoE dispatch/combine kernel.

\subsubsection{Computation--Communication Overlap}

\begin{table}[ht]
\small
\centering
\begin{tabular}{r rr rr}
\toprule
 & \multicolumn{2}{c}{Ours} & \multicolumn{2}{c}{\textit{pplx-kernels}} \\
\cmidrule(lr){2-3} \cmidrule(lr){4-5}
Batch & no overlap & dual-batch & no overlap & dual-batch \\
\midrule
128 & $11.813$ & $13.916$ & $1.548$ & $1.454$ \\
96  & $14.349$ & $16.487$ & $2.006$ & $1.807$ \\
64  & $21.260$ & $21.436$ & $2.824$ & $2.432$ \\
48  & $24.223$ & $24.197$ & $3.584$ & $3.272$ \\
32  & $32.003$ & $30.237$ & $4.903$ & $4.827$ \\
\bottomrule
\end{tabular}
\caption{End-to-end MoE decode speed (tokens/s) with and without dual-batch overlap (DeepSeek-V3, MTP, EP=DP=64, EFA)}
\label{table:dual-batch}
\end{table}

Dual-batch overlap pipelines computation of one batch with communication of another, and is effective only when the per-GPU batch size is sufficiently large.~\cite{Perplexity-DeepSeek, vLLM-DBO, SGLang-DBO}
\Cref{table:dual-batch} shows that even in throughput-oriented regimes, MoE dispatch/combine latency still matters: overlap provides only modest gains for our kernels and consistently degrades performance for \textit{pplx-kernels} due to their high communication latency.
For latency-sensitive workloads with small batches, computation and communication are sequential due to data dependencies, making low communication latency essential.

\begin{figure*}[t]
\centering
\includegraphics[width=\linewidth]{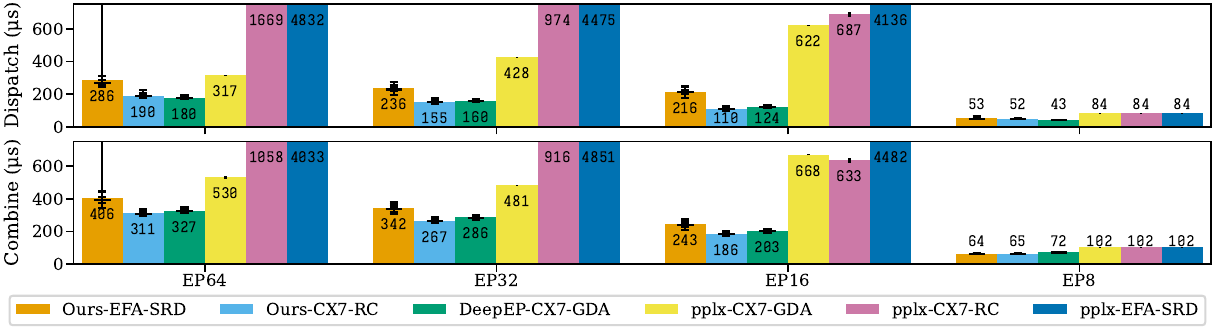}
\caption{MoE Decode Latency. Bar height is mean. Error bars show p01, p25, p50, p75, p95, p99. Error bars for \textit{pplx} indicate stddev.}
\label{fig:moe-decode}
\end{figure*}

\begin{figure*}[t]
\centering
\includegraphics[width=\linewidth]{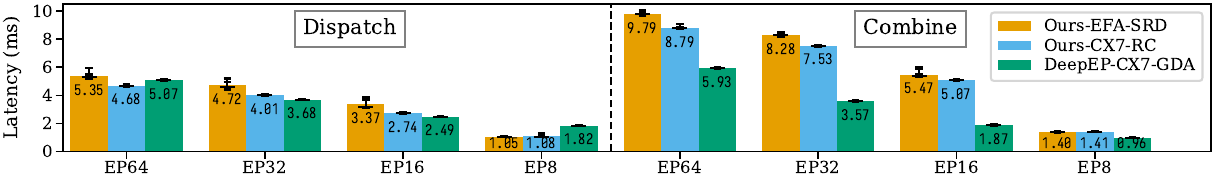}
\caption{MoE Prefill Latency. Bar height is mean. Error bars show p01, p25, p50, p75, p95, p99.}
\label{fig:moe-prefill}
\end{figure*}

\subsubsection{Kernel Microbenchmarks}

\paragraph{Microbenchmark Setup.}
To compare with DeepEP, our kernel-level benchmarks consider the settings of DeepSeek-V3/R1 model.
We dispatch $7168\times\texttt{fp8}$ tokens with $56\times\texttt{fp32}$ scaling factors and combine \texttt{bf16} tensors of the same dimension.
Decode is evaluated on batches of 128, prefill on chunks of 4096 tokens.
Each token is dispatched to 8 random experts.
We run 10,000 warmup iterations, followed by 10,000 benchmarked runs, aggregating timing across all ranks.
We simulate overlapped work and clear caches by inserting large GEMMs.

\paragraph{Decode Latency.}
\Cref{fig:moe-decode} shows the latencies of our MoE dispatch and combine kernels on decode-sized batches, comparing them to the specialized DeepEP and the portable \textit{pplx-kernels}.
Our proxy-based implementation not only makes EFA viable, it is also faster than the IBGDA-based kernels on ConnectX-7 and an order of magnitude faster than the NVSHMEM kernels using IBRC through the generic host proxy.

In the intra-node setup with 8 ranks, our kernels are slower by about \SI{2}{\micro\second} than DeepEP, largely due to the use of NICs to exchange routing information.
Whilst this data point is not critical for our primary inter-node use case, it does highlight that the NVLink transfers whose latencies are otherwise hidden by RDMA are highly efficient.
\textit{pplx-kernels} rely on fine-grained per-token NVLink synchronization under this configuration.
The discrepancy is due to bulk transfers and per-block/per-grid synchronization over NVLink.

Inter-node on 16 and 32 ranks our kernels outperform DeepEP on both dispatch and combine, largely due to the bulk transfers and efficient pipelining in the combine phase.
The ordering of RDMA and NVLink writes also helps reduce the latency to the first RDMA transfer.
When scaling to 64 ranks, combine still outperforms DeepEP, but the CPU overhead of the proxy thread becomes noticable and dispatch is slower due to the roughly microsecond overhead of enqueuing a transfer for each of the 56 inter-node peers.

Since the bandwidth is not saturated by decode, EFA latencies are trailing behind by only 30\%, despite 256KiB writes achieving less than half of the ConnectX-7 throughput.

\paragraph{Prefill Latency.}
\Cref{fig:moe-prefill} shows the latency of MoE Prefill.
We exclude \textit{pplx-kernels} from the comparison as they are not effective at 4096 tokens.
Due to the lack of chunking in transfers, the memory overhead of our decode-optimized kernels limits the set of models for which a deployment is viable.

On the dispatch side, DeepEP performs better with fewer ranks as they use RDMA to transfer only one replica of the token, moving copies of it to other ranks via NVLink.
At EP64, where the chances of multiple replicas landing on the same node are smaller, the effect of this optimization tapers off.
On combine, DeepEP's sender-side partial sum greatly reduces RDMA bytes, hence shows much lower latency, despite the reduced accumulation precision to \texttt{bf16}.

\subsubsection{Ablation: Private Buffer Size}

\begin{figure}[tb]
\centering
\includegraphics[width=\linewidth]{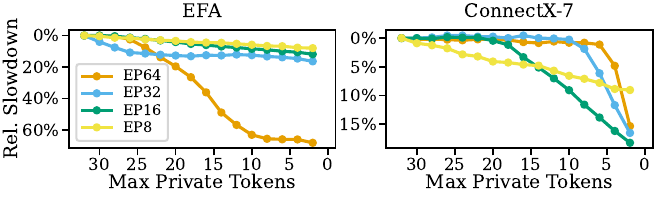}
\caption{Impact of private buffer size on p50 decode latency}
\label{fig:max-private-tokens}
\end{figure}

The private buffers were designed to hide the latency of routing information exchange.
We compare the median latency of total decode dispatch times while varying the number of tokens with the peak latency achieved with a buffer size that can accommodate all tokens in one burst.

\Cref{fig:max-private-tokens} indicates the performance dropoff as the private buffer size decreases.
The performance degradation in the absence of tokens justifies the use of the latency-reduction strategy.
In the intra-node case, where tokens are transferred faster, at least around 32 tokens are necessary to hide the latency of route exchange across both NICs.
In the inter-node case, ConnectX-7 NICs allow as few as 24 tokens to be used, while EFA NICs present a performance dropoff under 32 tokens already, since route exchange is slower.

\begin{table*}[t]
\small
\centering
\begin{tabular}{ll rrrr}
\toprule
Event & Thread & p50 & p90 & p99 & p99.9 \\
\midrule
\texttt{submit\_scatter()} & App & --- & --- & --- & --- \\
\textrightarrow\; Enqueue done & App & $0.120$ & $0.156$ & $2.019$ & $3.484$ \\
\textrightarrow\; Worker enqueue done & Worker & $0.855$ & $1.036$ & $1.450$ & $5.050$ \\
\textrightarrow\; Before posting first \textsc{Write} & Worker & $0.441$ & $0.535$ & $0.736$ & $4.889$ \\
\textrightarrow\; After posting last \textsc{Write} (EFA) & Worker & $27.886$ & $30.240$ & $39.176$ & $43.323$ \\
\quad\; After posting last \textsc{Write} (CX-7) & Worker & $8.502$ & $9.151$ & $12.310$ & $14.474$ \\
\bottomrule
\end{tabular}
\caption{CPU overhead breakdown for MoE all-to-all with EP64 (\SI{}{\micro\second})}
\label{table:cpu-overhead}
\end{table*}

\subsubsection{Ablation: Send and Receive Latency}

\begin{figure}[tb]
\centering
\includegraphics[width=\linewidth]{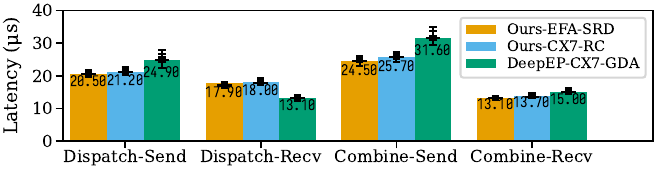}
\caption{Separate Send and Receive Latency for EP=64}
\label{fig:moe-send-recv}
\end{figure}

Our kernels, similarly to DeepEP, are split into senders and receivers, with both NVLink and RDMA transfers executing in the background between them.
\Cref{fig:moe-send-recv} shows these latencies, which were measured by inserting a long artificial delay simulating shared experts or overlapped work before the receive kernels, allowing all transfers to settle.

Dispatch and combine send outperform DeepEP as ours only copy memory.
Due to faster accumulation, combine receive is also faster.
Dispatch receive is an outlier, since it pulls data using NVLink loads.
Overall, the total execution time of these kernels is under 15\% of the transfer times.
More in-depth profiling reveals the proxy commences RDMA work midway through the execution of the send kernels, with shuffling adding only about \SI{15}{\micro\second} of idle time.

\subsubsection{Ablation: Host-Proxy Overhead Breakdown}

\begin{table}[ht]
\small
\centering
\begin{tabular}{ll rrrr}
\toprule
NIC & EP & p50 & p90 & p99 & p99.9 \\
\midrule
\multirow{4}{*}{EFA} & EP8  & $3.081$  & $3.348$  & $4.769$  & $13.175$ \\
 & EP16 & $6.536$  & $7.141$  & $8.429$  & $17.067$ \\
 & EP32 & $13.374$ & $14.412$ & $22.633$ & $26.803$ \\
 & EP64 & $27.886$ & $30.240$ & $39.176$ & $43.323$ \\
\midrule
\multirow{4}{*}{CX-7} & EP8  & $0.842$ & $0.893$ & $1.162$  & $3.281$ \\
 & EP16 & $1.926$ & $2.049$ & $2.853$  & $5.516$ \\
 & EP32 & $4.140$ & $4.392$ & $5.830$  & $8.709$ \\
 & EP64 & $8.502$ & $9.151$ & $12.310$ & $14.474$ \\
\bottomrule
\end{tabular}
\caption{Post time for all \textsc{Write}s of scatter (\SI{}{\micro\second})}
\label{table:scatter-scaling}
\end{table}

\Cref{table:cpu-overhead} instruments key events during MoE kernels with EP64, reporting the elapsed time since the previous event.
\textit{TransferEngine} incurs minimal CPU overhead from the application call to the first RDMA \textsc{Write}: under \SI{1.5}{\micro\second} at p50.
The dominant cost is posting the \textsc{Write}s themselves, which includes overhead inside \texttt{libfabric} for EFA.

\Cref{table:scatter-scaling} shows how posting time scales with the number of scatter targets.
Posting time increases roughly linearly with expert parallelism, but remains acceptable: at EP64 on ConnectX-7, the total posting time is under \SI{10}{\micro\second} at p50, and on EFA under \SI{28}{\micro\second}.
As shown in \Cref{fig:moe-decode}, this overhead does not prevent our kernels from matching or exceeding DeepEP performance.

\section{Discussion}

\paragraph{GPU-Initiated Communication}
GPU-initiated RDMA (GDA) bypasses the host CPU to reduce latency.
GDA is not yet available on most cloud instances (AWS \verb|p5|/\verb|p5e|, eRDMA) and remains preliminary on \verb|p5en|; our host-proxy design serves this current-generation hardware while preserving portability.
For MoE, next-generation GPUs with wide NVLink domain (e.g., GB200 NVL72) shift communication off RDMA entirely.
For KvCache and RL weight transfers, RDMA latency is already hidden by computation, so our CPU-based approach frees the GPU without sacrificing end-to-end performance.

\paragraph{Supporting Additional NICs}
The \textit{TransferEngine} currently supports ConnectX and EFA.
Among Linux \texttt{rdma\-core} providers, only EFA diverges from the standard RC transport.
For RC-compatible NICs (e.g., eRDMA, Broadcom, AMD), the internal implementation would resemble the ConnectX path.
Porting to additional NICs requires per-hardware tuning, not a redesign of the library; application code built on the \textit{TransferEngine} API remains unchanged.

\section{Conclusion}

Existing RDMA solutions for LLM systems suffer from vendor lock-in, with no viable implementations on custom cloud hardware such as AWS EFA.
\texttt{fabric-lib} addresses this by identifying common functionality across heterogeneous RDMA hardware.
By layering a reliable abstraction without ordering guarantees over the underlying protocols, we transparently extended support to multiple RDMA NICs, with a particular focus on EFA and ConnectX.

We demonstrated this approach through three production systems: KvCache transfer for disaggregated inference, RL weight updates achieving 1.3 seconds for trillion-parameter models, and MoE dispatch/combine with state-of-the-art latency on ConnectX-7 and the first viable implementation compatible with AWS EFA.
\texttt{fabric-lib} enables portable point-to-point communication for modern LLM architectures, avoiding vendor lock-in while complementing collective libraries for cloud-native deployments.

\section*{Acknowledgements}

We thank NVIDIA for generously providing access to the ConnectX-7 hardware for our evaluations, alongside valuable insights towards maximizing performance on H200 GPUs.
We thank AWS for their insights and advice towards improving performance on EFA.

\newpage
\bibliography{reference}

\begin{thebibliography}{56}
\providecommand{\natexlab}[1]{#1}
\providecommand{\url}[1]{\texttt{#1}}
\expandafter\ifx\csname urlstyle\endcsname\relax
  \providecommand{\doi}[1]{doi: #1}\else
  \providecommand{\doi}{doi: \begingroup \urlstyle{rm}\Url}\fi

\bibitem[Agostini et~al.(2018)Agostini, Rossetti, and Potluri]{GPUDirectAsync}
Agostini, E., Rossetti, D., and Potluri, S.
\newblock {GPUDirect} async: Exploring {GPU} synchronous communication
  techniques for {InfiniBand} clusters.
\newblock \emph{J. Parallel Distributed Comput.}, 114:\penalty0 28--45, 2018.
\newblock \doi{10.1016/J.JPDC.2017.12.007}.
\newblock URL \url{https://doi.org/10.1016/j.jpdc.2017.12.007}.

\bibitem[Ainslie et~al.(2023)Ainslie, Lee-Thorp, De~Jong, Zemlyanskiy,
  Lebr{\'o}n, and Sanghai]{ainslie2023gqa}
Ainslie, J., Lee-Thorp, J., De~Jong, M., Zemlyanskiy, Y., Lebr{\'o}n, F., and
  Sanghai, S.
\newblock Gqa: Training generalized multi-query transformer models from
  multi-head checkpoints.
\newblock \emph{arXiv preprint arXiv:2305.13245}, 2023.

\bibitem[{Alibaba Cloud}(2025)]{eRDMA}
{Alibaba Cloud}.
\newblock Elastic compute service: {eRDMA}, 2025.
\newblock URL
  \url{https://www.alibabacloud.com/help/en/ecs/user-guide/elastic-rdma-erdma/}.

\bibitem[Chang et~al.(2024)Chang, Bao, Hou, Jiang, Zheng, Zhong, Zhang, Song,
  Jiang, Lin, Jin, and Liu]{flux}
Chang, L.-W., Bao, W., Hou, Q., Jiang, C., Zheng, N., Zhong, Y., Zhang, X.,
  Song, Z., Jiang, Z., Lin, H., Jin, X., and Liu, X.
\newblock Flux: Fast software-based communication overlap on gpus through
  kernel fusion, 2024.

\bibitem[{DeepSeek AI}(2025)]{3fs}
{DeepSeek AI}.
\newblock Fire-flyer file system (3fs).
\newblock \url{https://github.com/deepseek-ai/3FS}, 2025.
\newblock High-performance distributed file system for AI training and
  inference workloads.

\bibitem[DeepSeek-AI et~al.(2025)DeepSeek-AI, Liu, Feng, Xue, Wang, Wu, Lu,
  Zhao, Deng, Zhang, Ruan, Dai, Guo, Yang, Chen, Ji, Li, Lin, Dai, Luo, Hao,
  Chen, Li, Zhang, Bao, Xu, Wang, Zhang, Ding, Xin, Gao, Li, Qu, Cai, Liang,
  Guo, Ni, Li, Wang, Chen, Chen, Yuan, Qiu, Li, Song, Dong, Hu, Gao, Guan,
  Huang, Yu, Wang, Zhang, Xu, Xia, Zhao, Wang, Zhang, Li, Wang, Zhang, Zhang,
  Tang, Li, Tian, Huang, Wang, Zhang, Wang, Zhu, Chen, Du, Chen, Jin, Ge,
  Zhang, Pan, Wang, Xu, Zhang, Chen, Li, Lu, Zhou, Chen, Wu, Ye, Ye, Ma, Wang,
  Zhou, Yu, Zhou, Pan, Wang, Yun, Pei, Sun, Xiao, Zeng, Zhao, An, Liu, Liang,
  Gao, Yu, Zhang, Li, Jin, Wang, Bi, Liu, Wang, Shen, Chen, Zhang, Chen, Nie,
  Sun, Wang, Cheng, Liu, Xie, Liu, Yu, Song, Shan, Zhou, Yang, Li, Su, Lin, Li,
  Wang, Wei, Zhu, Zhang, Xu, Xu, Huang, Li, Zhao, Sun, Li, Wang, Yu, Zheng,
  Zhang, Shi, Xiong, He, Tang, Piao, Wang, Tan, Ma, Liu, Guo, Wu, Ou, Zhu,
  Wang, Gong, Zou, He, Zha, Xiong, Ma, Yan, Luo, You, Liu, Zhou, Wu, Ren, Ren,
  Sha, Fu, Xu, Huang, Zhang, Xie, Zhang, Hao, Gou, Ma, Yan, Shao, Xu, Wu,
  Zhang, Li, Gu, Zhu, Liu, Li, Xie, Song, Gao, and Pan]{DeepSeek-V3}
DeepSeek-AI, Liu, A., Feng, B., Xue, B., Wang, B., Wu, B., Lu, C., Zhao, C.,
  Deng, C., Zhang, C., Ruan, C., Dai, D., Guo, D., Yang, D., Chen, D., Ji, D.,
  Li, E., Lin, F., Dai, F., Luo, F., Hao, G., Chen, G., Li, G., Zhang, H., Bao,
  H., Xu, H., Wang, H., Zhang, H., Ding, H., Xin, H., Gao, H., Li, H., Qu, H.,
  Cai, J.~L., Liang, J., Guo, J., Ni, J., Li, J., Wang, J., Chen, J., Chen, J.,
  Yuan, J., Qiu, J., Li, J., Song, J., Dong, K., Hu, K., Gao, K., Guan, K.,
  Huang, K., Yu, K., Wang, L., Zhang, L., Xu, L., Xia, L., Zhao, L., Wang, L.,
  Zhang, L., Li, M., Wang, M., Zhang, M., Zhang, M., Tang, M., Li, M., Tian,
  N., Huang, P., Wang, P., Zhang, P., Wang, Q., Zhu, Q., Chen, Q., Du, Q.,
  Chen, R.~J., Jin, R.~L., Ge, R., Zhang, R., Pan, R., Wang, R., Xu, R., Zhang,
  R., Chen, R., Li, S.~S., Lu, S., Zhou, S., Chen, S., Wu, S., Ye, S., Ye, S.,
  Ma, S., Wang, S., Zhou, S., Yu, S., Zhou, S., Pan, S., Wang, T., Yun, T.,
  Pei, T., Sun, T., Xiao, W.~L., Zeng, W., Zhao, W., An, W., Liu, W., Liang,
  W., Gao, W., Yu, W., Zhang, W., Li, X.~Q., Jin, X., Wang, X., Bi, X., Liu,
  X., Wang, X., Shen, X., Chen, X., Zhang, X., Chen, X., Nie, X., Sun, X.,
  Wang, X., Cheng, X., Liu, X., Xie, X., Liu, X., Yu, X., Song, X., Shan, X.,
  Zhou, X., Yang, X., Li, X., Su, X., Lin, X., Li, Y.~K., Wang, Y.~Q., Wei,
  Y.~X., Zhu, Y.~X., Zhang, Y., Xu, Y., Xu, Y., Huang, Y., Li, Y., Zhao, Y.,
  Sun, Y., Li, Y., Wang, Y., Yu, Y., Zheng, Y., Zhang, Y., Shi, Y., Xiong, Y.,
  He, Y., Tang, Y., Piao, Y., Wang, Y., Tan, Y., Ma, Y., Liu, Y., Guo, Y., Wu,
  Y., Ou, Y., Zhu, Y., Wang, Y., Gong, Y., Zou, Y., He, Y., Zha, Y., Xiong, Y.,
  Ma, Y., Yan, Y., Luo, Y., You, Y., Liu, Y., Zhou, Y., Wu, Z.~F., Ren, Z.~Z.,
  Ren, Z., Sha, Z., Fu, Z., Xu, Z., Huang, Z., Zhang, Z., Xie, Z., Zhang, Z.,
  Hao, Z., Gou, Z., Ma, Z., Yan, Z., Shao, Z., Xu, Z., Wu, Z., Zhang, Z., Li,
  Z., Gu, Z., Zhu, Z., Liu, Z., Li, Z., Xie, Z., Song, Z., Gao, Z., and Pan, Z.
\newblock Deepseek-v3 technical report, 2025.
\newblock URL \url{https://arxiv.org/abs/2412.19437}.

\bibitem[Fu et~al.(2025)Fu, Gao, Shen, Zhu, Mei, He, Xu, Wei, Mei, Wang, Yang,
  Yuan, and Wu]{fu2025areal}
Fu, W., Gao, J., Shen, X., Zhu, C., Mei, Z., He, C., Xu, S., Wei, G., Mei, J.,
  Wang, J., Yang, T., Yuan, B., and Wu, Y.
\newblock Areal: A large-scale asynchronous reinforcement learning system for
  language reasoning, 2025.
\newblock URL \url{https://arxiv.org/abs/2505.24298}.

\bibitem[He(2025)]{slime-issue132}
He, B.
\newblock [perf] weight sync optimization in colocate mode, slime {PR} \#132.
\newblock \url{https://github.com/THUDM/slime/issues/132}, 2025.
\newblock Accessed: 2025-10-14.

\bibitem[He et~al.(2025)He, Zhu, and
  Li]{BiaoHe-Optimizing-Weight-Sync-in-slime}
He, B., Zhu, Z., and Li, J.
\newblock Efficient rl training - optimizing weight sync in slime.
\newblock \url{https://hebiao064.github.io/rl-weight-sync}, 2025.
\newblock Accessed: 2025-10-14.

\bibitem[Hu et~al.(2025{\natexlab{a}})Hu, Wu, Shen, Liu, Zhu, Wang, Jiang,
  Wang, Chen, Chen, Fang, Xianyu, Cao, Xu, and Liu]{OpenRLHF}
Hu, J., Wu, X., Shen, W., Liu, J.~K., Zhu, Z., Wang, W., Jiang, S., Wang, H.,
  Chen, H., Chen, B., Fang, W., Xianyu, Cao, Y., Xu, H., and Liu, Y.
\newblock Openrlhf: An easy-to-use, scalable and high-performance rlhf
  framework, 2025{\natexlab{a}}.
\newblock URL \url{https://arxiv.org/abs/2405.11143}.

\bibitem[Hu et~al.(2025{\natexlab{b}})Hu, Shen, Bonato, Jeaugey, Alexander,
  Spada, Dinan, Hammond, and Hoefler]{Demystifying-NCCL}
Hu, Z., Shen, S., Bonato, T., Jeaugey, S., Alexander, C., Spada, E., Dinan, J.,
  Hammond, J., and Hoefler, T.
\newblock Demystifying {NCCL}: An in-depth analysis of {GPU} communication
  protocols and algorithms, 2025{\natexlab{b}}.
\newblock URL \url{https://arxiv.org/abs/2507.04786}.

\bibitem[Huang et~al.(2025)Huang, Chadha, Kong, Gao, and
  Li]{Nemo-RL-Optimizing-Weight-Transfer}
Huang, G., Chadha, P., Kong, T., Gao, W., and Li, Z.
\newblock {NeMo}-{RL}: Journey of optimizing weight transfer in large moe
  models by 10x.
\newblock \url{https://github.com/NVIDIA-NeMo/RL/discussions/1189}, 2025.
\newblock Accessed: 2025-10-14.

\bibitem[Kalia et~al.(2016)Kalia, Kaminsky, and
  Andersen]{DBLP:conf/usenix/KaliaKA16}
Kalia, A., Kaminsky, M., and Andersen, D.~G.
\newblock Design guidelines for high performance {RDMA} systems.
\newblock In Gulati, A. and Weatherspoon, H. (eds.), \emph{Proceedings of the
  2016 {USENIX} Annual Technical Conference, {USENIX} {ATC} 2016, Denver, CO,
  USA, June 22-24, 2016}, pp.\  437--450. {USENIX} Association, 2016.
\newblock URL
  \url{https://www.usenix.org/conference/atc16/technical-sessions/presentation/kalia}.

\bibitem[{Kimi Team} et~al.(2025){Kimi Team}, Bai, Bao, Chen, Chen, Chen, Chen,
  Chen, Chen, Chen, Chen, Cui, Ding, Dong, Du, Du, Du, Du, Fan, Feng, Fu, Gao,
  Gao, Gao, Gao, Gu, Guan, Guo, Guo, Hu, Hao, He, He, He, Hong, Hu, Hu, Huang,
  Huang, Huang, Jiang, Jiang, Jin, Kang, Lai, Li, Li, Li, Li, Li, Li, Li, Li,
  Li, Lin, Lin, Lin, Liu, Liu, Liu, Liu, Liu, Liu, Liu, Liu, Liu, Liu, Liu,
  Liu, Liu, Liu, Liu, Lu, Lu, Ma, Ma, Ma, Mao, Mei, Men, Miao, Pan, Peng, Qin,
  Qu, Shang, Shi, Shi, Song, Su, Su, Sun, Sung, Tang, Tao, Teng, Wang, Wang,
  Wang, Wang, Wang, Wang, Wang, Wang, Wang, Wang, Wang, Wang, Wang, Wang, Wang,
  Wang, Wang, Wei, Wei, Wu, Wu, Wu, Xiao, Xie, Xiong, Xu, Xu, Xu, Xu, Xu, Xu,
  Xu, Xu, Xu, Xu, Yan, Yan, Yang, Yang, Yang, Yang, Yang, Yao, Yao, Ye, Ye,
  Yin, Yu, Yuan, Yuan, Yuan, Zhan, Zhang, Zhang, Zhang, Zhang, Zhang, Zhang,
  Zhang, Zhang, Zhang, Zhang, Zhang, Zhao, Zhao, Zheng, Zheng, Zhou, Zhou,
  Zhou, Zhu, Zhuang, and Zu]{Kimi-K2}
{Kimi Team}, Bai, Y., Bao, Y., Chen, G., Chen, J., Chen, N., Chen, R., Chen,
  Y., Chen, Y., Chen, Y., Chen, Z., Cui, J., Ding, H., Dong, M., Du, A., Du,
  C., Du, D., Du, Y., Fan, Y., Feng, Y., Fu, K., Gao, B., Gao, H., Gao, P.,
  Gao, T., Gu, X., Guan, L., Guo, H., Guo, J., Hu, H., Hao, X., He, T., He, W.,
  He, W., Hong, C., Hu, Y., Hu, Z., Huang, W., Huang, Z., Huang, Z., Jiang, T.,
  Jiang, Z., Jin, X., Kang, Y., Lai, G., Li, C., Li, F., Li, H., Li, M., Li,
  W., Li, Y., Li, Y., Li, Z., Li, Z., Lin, H., Lin, X., Lin, Z., Liu, C., Liu,
  C., Liu, H., Liu, J., Liu, J., Liu, L., Liu, S., Liu, T.~Y., Liu, T., Liu,
  W., Liu, Y., Liu, Y., Liu, Y., Liu, Y., Liu, Z., Lu, E., Lu, L., Ma, S., Ma,
  X., Ma, Y., Mao, S., Mei, J., Men, X., Miao, Y., Pan, S., Peng, Y., Qin, R.,
  Qu, B., Shang, Z., Shi, L., Shi, S., Song, F., Su, J., Su, Z., Sun, X., Sung,
  F., Tang, H., Tao, J., Teng, Q., Wang, C., Wang, D., Wang, F., Wang, H.,
  Wang, J., Wang, J., Wang, J., Wang, S., Wang, S., Wang, Y., Wang, Y., Wang,
  Y., Wang, Y., Wang, Y., Wang, Z., Wang, Z., Wang, Z., Wei, C., Wei, Q., Wu,
  W., Wu, X., Wu, Y., Xiao, C., Xie, X., Xiong, W., Xu, B., Xu, J., Xu, J., Xu,
  L.~H., Xu, L., Xu, S., Xu, W., Xu, X., Xu, Y., Xu, Z., Yan, J., Yan, Y.,
  Yang, X., Yang, Y., Yang, Z., Yang, Z., Yang, Z., Yao, H., Yao, X., Ye, W.,
  Ye, Z., Yin, B., Yu, L., Yuan, E., Yuan, H., Yuan, M., Zhan, H., Zhang, D.,
  Zhang, H., Zhang, W., Zhang, X., Zhang, Y., Zhang, Y., Zhang, Y., Zhang, Y.,
  Zhang, Y., Zhang, Y., Zhang, Z., Zhao, H., Zhao, Y., Zheng, H., Zheng, S.,
  Zhou, J., Zhou, X., Zhou, Z., Zhu, Z., Zhuang, W., and Zu, X.
\newblock Kimi k2: Open agentic intelligence, 2025.
\newblock URL \url{https://arxiv.org/abs/2507.20534}.

\bibitem[Kwon et~al.(2023)Kwon, Li, Zhuang, Sheng, Zheng, Yu, Gonzalez, Zhang,
  and Stoica]{PagedAttention}
Kwon, W., Li, Z., Zhuang, S., Sheng, Y., Zheng, L., Yu, C.~H., Gonzalez, J.,
  Zhang, H., and Stoica, I.
\newblock Efficient memory management for large language model serving with
  pagedattention.
\newblock In Flinn, J., Seltzer, M.~I., Druschel, P., Kaufmann, A., and Mace,
  J. (eds.), \emph{Proceedings of the 29th Symposium on Operating Systems
  Principles, {SOSP} 2023, Koblenz, Germany, October 23-26, 2023}, pp.\
  611--626. {ACM}, 2023.
\newblock \doi{10.1145/3600006.3613165}.
\newblock URL \url{https://doi.org/10.1145/3600006.3613165}.

\bibitem[Langer et~al.(2021)Langer, Howell, Potluri, Dinan, and Kraus]{NVSHMEM}
Langer, A., Howell, S., Potluri, S., Dinan, J., and Kraus, J.
\newblock Dynamic symmetric heap allocation in {NVSHMEM}.
\newblock In \emph{OpenSHMEM and Related Technologies. OpenSHMEM in the Era of
  Exascale and Smart Networks: 8th Workshop on OpenSHMEM and Related
  Technologies, OpenSHMEM 2021, Virtual Event, September 14–16, 2021, Revised
  Selected Papers}, pp.\  187–198, Berlin, Heidelberg, 2021. Springer-Verlag.
\newblock ISBN 978-3-031-04887-6.
\newblock \doi{10.1007/978-3-031-04888-3_12}.
\newblock URL \url{https://doi.org/10.1007/978-3-031-04888-3_12}.

\bibitem[Li et~al.(2020)Li, Zhao, Varma, Salpekar, Noordhuis, Li, Paszke,
  Smith, Vaughan, Damania, and Chintala]{torch.distributed}
Li, S., Zhao, Y., Varma, R., Salpekar, O., Noordhuis, P., Li, T., Paszke, A.,
  Smith, J., Vaughan, B., Damania, P., and Chintala, S.
\newblock {PyTorch Distributed}: Experiences on accelerating data parallel
  training.
\newblock \emph{Proc. {VLDB} Endow.}, 13\penalty0 (12):\penalty0 3005--3018,
  2020.
\newblock \doi{10.14778/3415478.3415530}.
\newblock URL \url{http://www.vldb.org/pvldb/vol13/p3005-li.pdf}.

\bibitem[Li(2025)]{NeMo-RL-PR1267}
Li, Z.
\newblock feat: refit refactoring with zmq and overlapping, {NeMo}-{RL} {PR}
  \#1267.
\newblock \url{https://github.com/NVIDIA-NeMo/RL/pull/1267}, 2025.
\newblock Accessed: 2025-10-14.

\bibitem[Licker et~al.(2025)Licker, Hu, Zaytsev, and Chen]{pplx-kernels}
Licker, N., Hu, K., Zaytsev, V., and Chen, L.
\newblock {pplx-kernels}: {Perplexity} {MoE} kernels.
\newblock \url{https://github.com/perplexityai/pplx-kernels}, 2025.

\bibitem[Liu et~al.(2024)Liu, Feng, Wang, Wang, Liu, Zhao, Dengr, Ruan, Dai,
  Guo, et~al.]{liu2024deepseek}
Liu, A., Feng, B., Wang, B., Wang, B., Liu, B., Zhao, C., Dengr, C., Ruan, C.,
  Dai, D., Guo, D., et~al.
\newblock Deepseek-v2: A strong, economical, and efficient mixture-of-experts
  language model.
\newblock \emph{arXiv preprint arXiv:2405.04434}, 2024.

\bibitem[Mao et~al.(2025)Mao, Zhou, Zhang, Cui, Chen, and Xu]{UCCL-EP}
Mao, Z., Zhou, Y., Zhang, Y., Cui, C., Chen, Z., and Xu, Z.
\newblock Previewing {UCCL}-{EP}: Flexible and efficient expert parallelism for
  cloud and beyond.
\newblock \url{https://uccl-project.github.io/posts/uccl-ep/}, 2025.

\bibitem[Mei et~al.(2025)Mei, Fu, Li, Wang, Zhang, and Wu]{mei2025real}
Mei, Z., Fu, W., Li, K., Wang, G., Zhang, H., and Wu, Y.
\newblock Real: Efficient rlhf training of large language models with parameter
  reallocation.
\newblock In \emph{Proceedings of the Eighth Conference on Machine Learning and
  Systems, MLSys 2025, Santa Clara, CA, USA, May 12-15, 2025}. mlsys.org, 2025.

\bibitem[{Moonshot AI}(2025)]{MoonshotAI-Checkpoint-Engine}
{Moonshot AI}.
\newblock How {Kimi} {K2} achieves efficient {RL} parameter updates.
\newblock \url{https://moonshotai.github.io/checkpoint-engine/}, 2025.
\newblock Accessed: 2025-10-14.

\bibitem[{MPI Forum}(2025)]{mpi50}
{MPI Forum}.
\newblock \emph{{MPI}: A Message-Passing Interface Standard Version 5.0}, June
  2025.
\newblock URL \url{https://www.mpi-forum.org/docs/mpi-5.0/mpi50-report.pdf}.

\bibitem[{NVIDIA}(2012)]{GPUDirectRDMA}
{NVIDIA}.
\newblock {GPUDirect RDMA}, 2012.
\newblock URL \url{https://docs.nvidia.com/cuda/gpudirect-rdma/}.

\bibitem[{NVIDIA}(2015)]{NCCL}
{NVIDIA}.
\newblock {NVIDIA} collective communication library ({NCCL}), 2015.
\newblock URL \url{https://developer.nvidia.com/nccl}.

\bibitem[{NVIDIA}(2023)]{TensorRT-LLM}
{NVIDIA}.
\newblock {TensorRT}-{LLM}.
\newblock \url{https://github.com/ai-dynamo/nixl}, 2023.

\bibitem[{NVIDIA}(2025)]{NIXL}
{NVIDIA}.
\newblock {NIXL}: {NVIDIA} inference xfer library.
\newblock \url{https://github.com/ai-dynamo/nixl}, 2025.

\bibitem[OFIWG(2014)]{libfabric}
OFIWG.
\newblock {libfabric: Open Fabrics Interfaces (OFI)}, 2014.
\newblock URL \url{https://github.com/ofiwg/libfabric}.

\bibitem[Patel et~al.(2024)Patel, Choukse, Zhang, Shah, Goiri, Maleki, and
  Bianchini]{splitwise}
Patel, P., Choukse, E., Zhang, C., Shah, A., Goiri, {\'{I}}., Maleki, S., and
  Bianchini, R.
\newblock Splitwise: Efficient generative {LLM} inference using phase
  splitting.
\newblock In \emph{51st {ACM/IEEE} Annual International Symposium on Computer
  Architecture, {ISCA} 2024, Buenos Aires, Argentina, June 29 - July 3, 2024},
  pp.\  118--132. {IEEE}, 2024.
\newblock \doi{10.1109/ISCA59077.2024.00019}.
\newblock URL \url{https://doi.org/10.1109/ISCA59077.2024.00019}.

\bibitem[{Perplexity AI}(2025)]{Perplexity-DeepSeek}
{Perplexity AI}.
\newblock Lower latency and higher throughput with multi-node {DeepSeek}
  deployment.
\newblock
  \url{https://research.perplexity.ai/articles/lower-latency-and-higher-throughput-with-multi-node-deepseek-deployment},
  2025.

\bibitem[Qin et~al.(2025)Qin, Li, He, Cui, Ren, Zhang, Wu, Zheng, and
  Xu]{Mooncake2025}
Qin, R., Li, Z., He, W., Cui, J., Ren, F., Zhang, M., Wu, Y., Zheng, W., and
  Xu, X.
\newblock {Mooncake}: Trading more storage for less computation {\textemdash} a
  {KVCache-centric} architecture for serving {LLM} chatbot.
\newblock In \emph{23rd USENIX Conference on File and Storage Technologies
  (FAST 25)}, pp.\  155--170, Santa Clara, CA, February 2025. USENIX
  Association.
\newblock ISBN 978-1-939133-45-8.
\newblock URL \url{https://www.usenix.org/conference/fast25/presentation/qin}.

\bibitem[Reda et~al.(2022)Reda, Canini, Kostic, and
  Peter]{DBLP:conf/nsdi/RedaCKP22}
Reda, W., Canini, M., Kostic, D., and Peter, S.
\newblock {RDMA} is turing complete, we just did not know it yet!
\newblock In Phanishayee, A. and Sekar, V. (eds.), \emph{19th {USENIX}
  Symposium on Networked Systems Design and Implementation, {NSDI} 2022,
  Renton, WA, USA, April 4-6, 2022}, pp.\  71--85. {USENIX} Association, 2022.
\newblock URL \url{https://www.usenix.org/conference/nsdi22/presentation/reda}.

\bibitem[Sergeev \& Balso(2018)Sergeev and Balso]{Horovod}
Sergeev, A. and Balso, M.~D.
\newblock {Horovod}: fast and easy distributed deep learning in tensorflow,
  2018.
\newblock URL \url{https://arxiv.org/abs/1802.05799}.

\bibitem[{SGLang}(2025)]{SGLang-DBO}
{SGLang}.
\newblock Deploying {DeepSeek} with {PD} disaggregation and large-scale expert
  parallelism on 96 {H100} {GPUs}.
\newblock \url{https://lmsys.org/blog/2025-05-05-large-scale-ep/}, 2025.

\bibitem[Shah et~al.(2025)Shah, Jangda, Li, Rocha, Hwang, Jose, Musuvathi,
  Saarikivi, Cheng, Zhou, Dathathri, Maleki, and Yang]{MSCCLpp}
Shah, A., Jangda, A., Li, B., Rocha, C., Hwang, C., Jose, J., Musuvathi, M.,
  Saarikivi, O., Cheng, P., Zhou, Q., Dathathri, R., Maleki, S., and Yang, Z.
\newblock Msccl++: Rethinking gpu communication abstractions for cutting-edge
  ai applications, 2025.
\newblock URL \url{https://arxiv.org/abs/2504.09014}.

\bibitem[Shalev et~al.(2020)Shalev, Ayoub, Bshara, and Sabbag]{AWS-SRD}
Shalev, L., Ayoub, H., Bshara, N., and Sabbag, E.
\newblock A cloud-optimized transport protocol for elastic and scalable {HPC}.
\newblock \emph{{IEEE} Micro}, 40\penalty0 (6):\penalty0 67--73, 2020.
\newblock \doi{10.1109/MM.2020.3016891}.
\newblock URL \url{https://doi.org/10.1109/MM.2020.3016891}.

\bibitem[Shamis et~al.(2015)Shamis, Venkata, Lopez, Baker, Hernandez, Itigin,
  Dubman, Shainer, Graham, Liss, et~al.]{UCX}
Shamis, P., Venkata, M.~G., Lopez, M.~G., Baker, M.~B., Hernandez, O., Itigin,
  Y., Dubman, M., Shainer, G., Graham, R.~L., Liss, L., et~al.
\newblock {UCX}: an open source framework for {HPC} network {API}s and beyond.
\newblock In \emph{2015 IEEE 23rd Annual Symposium on High-Performance
  Interconnects}, pp.\  40--43. IEEE, 2015.

\bibitem[Shazeer et~al.(2017)Shazeer, Mirhoseini, Maziarz, Davis, Le, Hinton,
  and Dean]{DBLP:conf/iclr/ShazeerMMDLHD17}
Shazeer, N., Mirhoseini, A., Maziarz, K., Davis, A., Le, Q.~V., Hinton, G.~E.,
  and Dean, J.
\newblock Outrageously large neural networks: The sparsely-gated
  mixture-of-experts layer.
\newblock In \emph{5th International Conference on Learning Representations,
  {ICLR} 2017, Toulon, France, April 24-26, 2017, Conference Track
  Proceedings}. OpenReview.net, 2017.
\newblock URL \url{https://openreview.net/forum?id=B1ckMDqlg}.

\bibitem[Sheng et~al.(2024)Sheng, Zhang, Ye, Wu, Zhang, Zhang, Peng, Lin, and
  Wu]{HybridFlow}
Sheng, G., Zhang, C., Ye, Z., Wu, X., Zhang, W., Zhang, R., Peng, Y., Lin, H.,
  and Wu, C.
\newblock Hybridflow: A flexible and efficient rlhf framework.
\newblock \emph{arXiv preprint arXiv: 2409.19256}, 2024.

\bibitem[Shi et~al.(2014)Shi, Potluri, Hamidouche, Perkins, Li, Rossetti, and
  Panda]{GDRCopy}
Shi, R., Potluri, S., Hamidouche, K., Perkins, J.~L., Li, M., Rossetti, D., and
  Panda, D.~K.
\newblock Designing efficient small message transfer mechanism for inter-node
  {MPI} communication on {InfiniBand} {GPU} clusters.
\newblock In \emph{21st International Conference on High Performance Computing,
  HiPC 2014, Goa, India, December 17-20, 2014}, pp.\  1--10. {IEEE} Computer
  Society, 2014.
\newblock \doi{10.1109/HIPC.2014.7116873}.
\newblock URL \url{https://doi.org/10.1109/HiPC.2014.7116873}.

\bibitem[Shoeybi et~al.(2020)Shoeybi, Patwary, Puri, LeGresley, Casper, and
  Catanzaro]{Megatron-LM}
Shoeybi, M., Patwary, M., Puri, R., LeGresley, P., Casper, J., and Catanzaro,
  B.
\newblock {Megatron-LM}: Training multi-billion parameter language models using
  model parallelism, 2020.
\newblock URL \url{https://arxiv.org/abs/1909.08053}.

\bibitem[Singhvi et~al.(2025)Singhvi, Dukkipati, Chandra, Wassel, Sharma,
  Rebello, Schuh, Kumar, Montazeri, Bansod, Thomas, Cho, Seibert, Wu, Yang, Li,
  Huang, Yin, Agarwal, Vaduvatha, Wang, Moshref, Ji, Wetherall, and
  Vahdat]{Falcon}
Singhvi, A., Dukkipati, N., Chandra, P., Wassel, H. M.~G., Sharma, N.~K.,
  Rebello, A., Schuh, H., Kumar, P., Montazeri, B., Bansod, N., Thomas, S.,
  Cho, I., Seibert, H.~L., Wu, B., Yang, R., Li, Y., Huang, K., Yin, Q.,
  Agarwal, A., Vaduvatha, S., Wang, W., Moshref, M., Ji, T., Wetherall, D., and
  Vahdat, A.
\newblock Falcon: {A} reliable, low latency hardware transport.
\newblock In Curado, M., Rothenberg, C.~E., Porter, G., and Kandula, S. (eds.),
  \emph{Proceedings of the {ACM} {SIGCOMM} 2025 Conference, {SIGCOMM} 2025,
  S{\~{a}}o Francisco Convent, Coimbra, Portugal, September 8-11, 2025}, pp.\
  248--263. {ACM}, 2025.
\newblock \doi{10.1145/3718958.3754353}.
\newblock URL \url{https://doi.org/10.1145/3718958.3754353}.

\bibitem[{vLLM}(2025)]{vLLM-DBO}
{vLLM}.
\newblock Dual batch overlap.
\newblock
  \url{https://github.com/vllm-project/vllm/blob/v0.13.0/docs/design/dbo.md},
  2025.

\bibitem[Wu et~al.(2025)Wu, Wang, Tang, Ding, Helenowski, Tan, Xu, Gowda, Chen,
  Zhu, Tang, Qian, Zhu, and Hou]{LlamaRL}
Wu, B., Wang, S., Tang, Y., Ding, J., Helenowski, E., Tan, L., Xu, T., Gowda,
  T., Chen, Z., Zhu, C., Tang, X., Qian, Y., Zhu, B., and Hou, R.
\newblock {LlamaRL}: A distributed asynchronous reinforcement learning
  framework for efficient large-scale {LLM} training, 2025.
\newblock URL \url{https://arxiv.org/abs/2505.24034}.

\bibitem[Yang et~al.(2025)Yang, Li, Yang, Zhang, Hui, Zheng, Yu, Gao, Huang,
  Lv, Zheng, Liu, Zhou, Huang, Hu, Ge, Wei, Lin, Tang, Yang, Tu, Zhang, Yang,
  Yang, Zhou, Zhou, Lin, Dang, Bao, Yang, Yu, Deng, Li, Xue, Li, Zhang, Wang,
  Zhu, Men, Gao, Liu, Luo, Li, Tang, Yin, Ren, Wang, Zhang, Ren, Fan, Su,
  Zhang, Zhang, Wan, Liu, Wang, Cui, Zhang, Zhou, and Qiu]{Qwen3}
Yang, A., Li, A., Yang, B., Zhang, B., Hui, B., Zheng, B., Yu, B., Gao, C.,
  Huang, C., Lv, C., Zheng, C., Liu, D., Zhou, F., Huang, F., Hu, F., Ge, H.,
  Wei, H., Lin, H., Tang, J., Yang, J., Tu, J., Zhang, J., Yang, J., Yang, J.,
  Zhou, J., Zhou, J., Lin, J., Dang, K., Bao, K., Yang, K., Yu, L., Deng, L.,
  Li, M., Xue, M., Li, M., Zhang, P., Wang, P., Zhu, Q., Men, R., Gao, R., Liu,
  S., Luo, S., Li, T., Tang, T., Yin, W., Ren, X., Wang, X., Zhang, X., Ren,
  X., Fan, Y., Su, Y., Zhang, Y., Zhang, Y., Wan, Y., Liu, Y., Wang, Z., Cui,
  Z., Zhang, Z., Zhou, Z., and Qiu, Z.
\newblock Qwen3 technical report, 2025.
\newblock URL \url{https://arxiv.org/abs/2505.09388}.

\bibitem[Ye et~al.(2025)Ye, Chen, Lai, Lin, Zhang, Wang, Chen, Kasikci, Grover,
  Krishnamurthy, and Ceze]{FlashInfer}
Ye, Z., Chen, L., Lai, R., Lin, W., Zhang, Y., Wang, S., Chen, T., Kasikci, B.,
  Grover, V., Krishnamurthy, A., and Ceze, L.
\newblock {FlashInfer}: Efficient and customizable attention engine for {LLM}
  inference serving, 2025.
\newblock URL \url{https://arxiv.org/abs/2501.01005}.

\bibitem[Zhang et~al.(2025)Zhang, Zheng, Lin, Jiang, Bao, Jiang, Hou, Cui,
  Zheng, Chang, Chen, and Liu]{comet}
Zhang, S., Zheng, N., Lin, H., Jiang, Z., Bao, W., Jiang, C., Hou, Q., Cui, W.,
  Zheng, S., Chang, L.-W., Chen, Q., and Liu, X.
\newblock {COMET}: Fine-grained computation-communication overlapping for
  mixture-of-experts.
\newblock In \emph{Eighth Conference on Machine Learning and Systems}, 2025.
\newblock URL \url{https://openreview.net/forum?id=fGgQS5VW09}.

\bibitem[Zhao et~al.(2025)Zhao, Zhou, Zhang, Deng, Xu, Liu, Yu, Li, and
  Zhao]{DeepEP}
Zhao, C., Zhou, S., Zhang, L., Deng, C., Xu, Z., Liu, Y., Yu, K., Li, J., and
  Zhao, L.
\newblock {DeepEP}: an efficient expert-parallel communication library.
\newblock \url{https://github.com/deepseek-ai/DeepEP}, 2025.

\bibitem[Zhao et~al.(2023)Zhao, Gu, Varma, Luo, Huang, Xu, Wright, Shojanazeri,
  Ott, Shleifer, Desmaison, Balioglu, Damania, Nguyen, Chauhan, Hao, Mathews,
  and Li]{FSDP}
Zhao, Y., Gu, A., Varma, R., Luo, L., Huang, C.-C., Xu, M., Wright, L.,
  Shojanazeri, H., Ott, M., Shleifer, S., Desmaison, A., Balioglu, C., Damania,
  P., Nguyen, B., Chauhan, G., Hao, Y., Mathews, A., and Li, S.
\newblock {PyTorch FSDP}: Experiences on scaling fully sharded data parallel.
\newblock \emph{Proc. VLDB Endow.}, 16\penalty0 (12):\penalty0 3848--3860,
  August 2023.
\newblock ISSN 2150-8097.
\newblock \doi{10.14778/3611540.3611569}.
\newblock URL \url{https://doi.org/10.14778/3611540.3611569}.

\bibitem[Zheng et~al.(2024)Zheng, Yin, Xie, Sun, Huang, Yu, Cao, Kozyrakis,
  Stoica, Gonzalez, Barrett, and Sheng]{SGLang}
Zheng, L., Yin, L., Xie, Z., Sun, C., Huang, J., Yu, C.~H., Cao, S., Kozyrakis,
  C., Stoica, I., Gonzalez, J.~E., Barrett, C.~W., and Sheng, Y.
\newblock Sglang: Efficient execution of structured language model programs.
\newblock In Globersons, A., Mackey, L., Belgrave, D., Fan, A., Paquet, U.,
  Tomczak, J.~M., and Zhang, C. (eds.), \emph{Advances in Neural Information
  Processing Systems 38: Annual Conference on Neural Information Processing
  Systems 2024, NeurIPS 2024, Vancouver, BC, Canada, December 10 - 15, 2024},
  2024.
\newblock URL
  \url{http://papers.nips.cc/paper\_files/paper/2024/hash/724be4472168f31ba1c9ac630f15dec8-Abstract-Conference.html}.

\bibitem[Zheng et~al.(2025{\natexlab{a}})Zheng, Bao, Hou, Zheng, Fang, Huang,
  Li, Duanmu, Chen, Xu, Guo, Zheng, Jiang, Di, Wang, Ye, Lin, Chang, Lu, Liang,
  Zhai, and Liu]{Triton-distributed}
Zheng, S., Bao, W., Hou, Q., Zheng, X., Fang, J., Huang, C., Li, T., Duanmu,
  H., Chen, R., Xu, R., Guo, Y., Zheng, N., Jiang, Z., Di, X., Wang, D., Ye,
  J., Lin, H., Chang, L.-W., Lu, L., Liang, Y., Zhai, J., and Liu, X.
\newblock Triton-distributed: Programming overlapping kernels on distributed ai
  systems with the triton compiler, 2025{\natexlab{a}}.
\newblock URL \url{https://arxiv.org/abs/2504.19442}.

\bibitem[Zheng et~al.(2025{\natexlab{b}})Zheng, Fang, Zheng, Hou, Bao, Zheng,
  Jiang, Wang, Ye, Lin, et~al.]{Tilelink}
Zheng, S., Fang, J., Zheng, X., Hou, Q., Bao, W., Zheng, N., Jiang, Z., Wang,
  D., Ye, J., Lin, H., et~al.
\newblock Tilelink: Generating efficient compute-communication overlapping
  kernels using tile-centric primitives.
\newblock \emph{arXiv preprint arXiv:2503.20313}, 2025{\natexlab{b}}.

\bibitem[Zhong et~al.(2024)Zhong, Liu, Chen, Hu, Zhu, Liu, Jin, and
  Zhang]{DistServe}
Zhong, Y., Liu, S., Chen, J., Hu, J., Zhu, Y., Liu, X., Jin, X., and Zhang, H.
\newblock Distserve: Disaggregating prefill and decoding for goodput-optimized
  large language model serving.
\newblock In Gavrilovska, A. and Terry, D.~B. (eds.), \emph{18th {USENIX}
  Symposium on Operating Systems Design and Implementation, {OSDI} 2024, Santa
  Clara, CA, USA, July 10-12, 2024}, pp.\  193--210. {USENIX} Association,
  2024.
\newblock URL
  \url{https://www.usenix.org/conference/osdi24/presentation/zhong-yinmin}.

\bibitem[Zhou et~al.(2025)Zhou, Chen, Mao, Lao, Yang, Kannan, Gao, Zhao, Wu,
  You, Ren, Xu, Raiciu, and Stoica]{UCCL}
Zhou, Y., Chen, Z., Mao, Z., Lao, C., Yang, S., Kannan, P.~G., Gao, J., Zhao,
  Y., Wu, Y., You, K., Ren, F., Xu, Z., Raiciu, C., and Stoica, I.
\newblock An extensible software transport layer for gpu networking, 2025.
\newblock URL \url{https://arxiv.org/abs/2504.17307}.

\bibitem[Zhu et~al.(2025)Zhu, Xie, Lv, and slime Contributors]{slime_github}
Zhu, Z., Xie, C., Lv, X., and slime Contributors.
\newblock slime: An llm post-training framework for rl scaling.
\newblock \url{https://github.com/THUDM/slime}, 2025.
\newblock GitHub repository. Corresponding author: Xin Lv.

\end{thebibliography}
\bibliographystyle{mlsys2025}

\newpage
\appendix
\section{KvCache Transfer Pseudocode}
\label{sec:appendix-kvcache}

\begin{figure}[h]
\begin{minted}[baselinestretch=1,fontsize=\fontsize{8pt}{8pt}\selectfont]{rust}
struct DispatchReq {
  input_ids: Vec<u32>,
  decoder_addr: NetAddr, imm: u32,
  kv_desc: MrDesc, pages: Vec<u32>,
  tail_desc: MrDesc, tail_idx: u32,
}
\end{minted}
\caption{Message sent from decoder to prefiller via \textsc{Send}/\textsc{Recv}.}
\label{code:kv-dispatch-req}
\end{figure}

\begin{figure}[h]
\begin{minted}[baselinestretch=1,fontsize=\fontsize{8pt}{8pt}\selectfont]{python}
def decoder_engine(req: Request, prefiller: NetAddr):
  # Allocate from GPU memory pool
  page_indices: list[int] = alloc_pages(req)
  tail_idx: int = alloc_tail()
  # Set up ImmCounter callback
  imm = alloc_imm()
  imm_count = len(page_indices) * n_layers + 1
  ev = Event()
  te.expect_imm_count(imm, imm_count, lambda: ev.set())
  # Dispatch to prefiller via RDMA SEND
  te.submit_send(prefiller, serialize(DispatchReq(...)))
  # Start decode once received all WRITEs from Prefiller
  ev.wait()
  process_tail(tail[tail_idx])
  free_imm(imm); free_tail(tail_idx)
  auto_regressive_decode(...)
  free_pages(...)
\end{minted}
\caption{Decoder workflow}
\label{code:kv-decoder}
\end{figure}

\begin{figure}[h]
\begin{minted}[baselinestretch=1,fontsize=\fontsize{8pt}{8pt}\selectfont]{python}
def prefiller_init():
  te.submit_recvs(MAX_LEN, MAX_RECV,
    lambda msg: prefiller_engine(deserialize(msg)))

def prefiller_engine(req: DispatchReq):
  # Allocate from GPU memory pool
  page_indices: list[int] = alloc_pages(req)
  tail_idx: int = alloc_tail()
  # Set up UvmWatcher (CUDA graph compatible)
  cnt_done = 0
  def uvm_cb(old: int, new: int):
    for layer in range(old, new):
      # RDMA WRITE pages to decoder
      te.submit_paged_writes(...,
        on_done=lambda: cnt_done += n_pages)
  watcher_ptr = te.alloc_uvm_watcher(uvm_cb)
  # Model forward
  x = embed(req.input_ids)
  for i in range(num_layers):
    x = self_attn(x)
    scalar_inc_(watcher_ptr)
    x = mlp(x)
  x = lm_head(x)
  tail[tail_idx].copy_(x) # Or other tail context
  # RDMA WRITE tail context to decoder
  tail[tail_idx].synchronize()
  te.submit_single_write(...,
    on_done=lambda: cnt_done += 1)
  # Wait until all RDMA WRITEs are done
  while cnt_done != n_pages * n_layers + 1:
    sleep(0)
  free_tail(tail_idx); free_pages(page_indices)
\end{minted}
\caption{Prefiller workflow}
\label{code:kv-prefiller}
\end{figure}

This section provides pseudocode for the disaggregated KvCache transfer workflow described in \Cref{sec:kvcache}, using the \textit{TransferEngine} API from \Cref{code:api}.
For clarity, the pseudocode uses a synchronous, blocking style; in practice, all operations are asynchronous tasks on a Tokio event loop.
\emph{Tail context} refers to auxiliary GPU tensors transferred from prefiller to decoder beyond the KvCache---in this example, the output logits.
The specific contents are inference-engine-dependent, but the transfer pattern is the same.

For brevity, the pseudocode omits:
(1)~sharding and replication of KV heads for tensor parallelism and Multi-head Latent Attention (MLA);
(2)~support for multiple KvCache buffers (e.g., separate key and value buffers);
(3)~chunked prefill, which does not change the total number of transferred pages or the expected \texttt{imm\_count} but requires additional bookkeeping;
(4)~prefix-cache reuse, where pages already present on the decoder need not be transferred from the prefiller;
(5)~stride and offset computations for \textsc{Write}s.

\Cref{code:kv-dispatch-req} shows the message sent from decoder to prefiller.
The decoder includes its RDMA memory descriptors (\texttt{kv\_desc}, \texttt{tail\_desc}) and page indices so the prefiller can \textsc{Write} directly into the decoder's GPU memory.

\Cref{code:kv-decoder} shows the decoder workflow.
The decoder pre-registers an \textsc{ImmCounter} with the expected number of \textsc{WriteImm} completions before dispatching, so no explicit completion message is needed from the prefiller.

\Cref{code:kv-prefiller} shows the prefiller workflow.
UVM watcher overlaps KvCache transfer with computation. After each layer's attention, the GPU increments a UVM counter via \texttt{scalar\_inc\_}; the CPU-side callback fires and submits \textsc{WriteImm} for that layer while the next layer computes.

\section{RL Weight Transfer Pseudocode}
\label{sec:appendix-rl}

\begin{figure}[h]
\begin{minted}[baselinestretch=1,fontsize=\fontsize{8pt}{8pt}\selectfont]{python}
@ray.remote
class TrainingWorker:
  def param_meta(self) -> dict[str, ParamMeta]: ...
  def set_routing(self, route: RoutingTable) -> None: ...
  def transfer_weights(self) -> None: ...
@ray.remote
class RolloutWorker:
  def param_meta(self) -> dict[str, ParamMeta]: ...
  def memory_regions(self) -> list[MemoryRegion]: ...
def controller_main() -> None:
  # Init and gather metadata
  trainers: list[TrainingWorker] = ...
  rollouts: list[Rolloutworker] = ...
  t_params = ray.get([x.param_meta() for x in trainers])
  r_params = ray.get([x.param_meta() for x in rollouts])
  r_mrs = ray.get([x.memory_regions() for x in rollouts])
  # Compute and set weight transfer routing
  route = compute_routing(t_params, r_params, r_mrs)
  ray.get([x.set_routing(route) for x in trainers])
  # Training loop
  while training_not_done:
    train()
    ray.get([x.transfer_weights() for x in trainers])
    rollout()
\end{minted}
\caption{RL training and weight update workflow}
\label{code:rl-flow}
\end{figure}

\Cref{code:rl-flow} shows the RL weight transfer workflow described in \Cref{sec:rollout}.
The controller gathers parameter metadata and memory region descriptors from all workers, computes a static routing table that maps each training GPU's parameters to the target inference GPU memory regions, and broadcasts the routing to training workers.
At each training step, training workers execute the weight transfer using the pre-computed routing, without re-planning or coordination.

\end{document}